\documentclass[acmtog,screen,nonacm]{acmart}

\usepackage{multirow}
\usepackage{amsmath}
\usepackage{amssymb}
\usepackage{enumitem}
\usepackage{layouts}
\usepackage{hyphenat,balance,microtype}
\usepackage[fleqn,tbtags]{mathtools}
\usepackage[capitalise]{cleveref}
\usepackage[detect-weight]{siunitx}
\usepackage{caption,subcaption,textpos}
\usepackage{colortbl}

\usepackage{algorithm}
\usepackage{algorithmic}
\usepackage{verbatim}
\usepackage{listings}
\usepackage{wrapfig}

\graphicspath{{./}} 

\usepackage{booktabs}
\usepackage{tabularx}
\usepackage{array}
\usepackage{fvextra}
\usepackage{xcolor}
\usepackage{listings}
\usepackage[detect-weight]{siunitx}

\definecolor{promptbg}{HTML}{FBFBFB}

\newcommand{\rh}[1]{#1}

\newcommand{\tablestyle}[2]{\setlength{\tabcolsep}{#1}
                            \renewcommand{\arraystretch}{#2}
                            \centering
                            \footnotesize} 
\definecolor{graycolor}{gray}{.9}

\title{Agentic 3D Creation via Joint Agent-Program Design}

\author{Rui-Huan Wang}
\email{2501112180@stu.pku.edu.cn}
\affiliation{
 \institution{Peking University}
 \country{China}
}

\author{Si-Tong Wei}
\email{weisitong@pku.edu.cn}
\affiliation{
 \institution{Peking University}
 \country{China}
}

\author{Jia-Qi He}
\email{hejiaqi2024@hotmail.com}
\affiliation{
 \institution{Peking University}
 \country{China}
}

\author{Heng-Yi Wei}
\email{2300013223@stu.pku.edu.cn}
\affiliation{
 \institution{Peking University}
 \country{China}
}

\author{Baoquan Chen}
\email{baoquan@pku.edu.cn}
\affiliation{
 \institution{Peking University}
 \country{China}
}

\author{Peng-Shuai Wang}
\email{wangps@hotmail.com}
\authornote{Corresponding author.}
\affiliation{
 \institution{Peking University}
 \country{China}
}

\begin{abstract}
Programmatic representations provide a compelling paradigm for 3D content creation, enabling fine-grained edits, interpretability, and explicit structural control.
Yet, agentic workflows that rely on large language models (LLMs) to author 3D programs remain brittle, often failing to translate high-level intent into consistent low-level geometry.
We attribute this fragility to a mismatch between existing programmatic interfaces and the reasoning strengths of LLMs, which favor semantic structure and spatial relations over fragile numeric choices.
In this paper, we jointly design an Agent-centric Domain-Specific Language (\textsc{aDSL}) and a role-specialized multi-agent system to close this gap.
\textsc{aDSL} bridges semantic logic and geometric constraints by emphasizing composability and spatial reasoning; it enables agents to manipulate geometry through relational operators instead of brittle absolute coordinates.
Building on \textsc{aDSL}, our training-free multi-agent system follows a Plan--Execute--Critic loop to decompose requests, synthesize code, and iteratively repair errors and constraint violations using execution feedback.
Experiments show that this co-design improves robustness, controllability, and faithfulness to user intent.
Our method outperforms prior LLM-based baselines on text-to-shape and image-to-shape tasks while preserving explicit structure, editability, and interpretability. It also enables downstream applications such as articulated object creation and structured scene composition.
\emph{Our code is available at \url{https://github.com/sig-pku/aDSL}.}
\end{abstract}

\begin{teaserfigure}
  \centering
  \includegraphics[width=\linewidth]{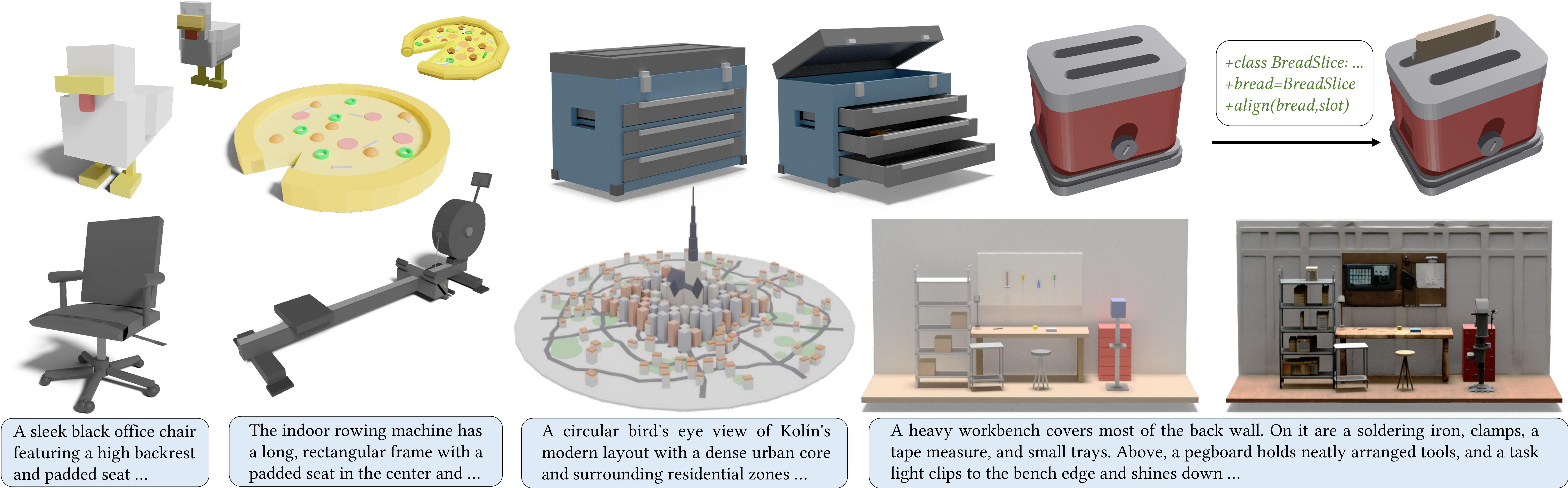}
  \caption{We present a jointly designed DSL and multi-agent system for robust text- and image-conditioned 3D modeling. The same structured representation also supports downstream applications including articulated asset modeling, shape editing and scene-level creation.}
  \label{fig:teaser}
\end{teaserfigure}

\begin{document}

\maketitle
\section{Introduction} \label{sec:intro}


3D content creation is a foundational problem in computer graphics, supporting applications spanning games, film production, and embodied AI.
Recent advances in 3D generation have been largely driven by diffusion models trained on large-scale 3D datasets and diverse geometric representations~\cite{Xiong2024,Xiang2024,He2025,Hunyuan3d2025}.
In parallel, an alternative paradigm represents 3D shapes and scenes as \emph{programs} and leverages LLM-based agents to generate and manipulate them~\cite{lu2025ll3m,du2024blenderllm,zhang2025shapecraft,hu2024scenecraft,zhang2025scene}.
By explicitly encoding structure, parameters, and design constraints, programmatic representations enable an \emph{agentic} workflow with better controllability, interpretability, and editability. \looseness=-1

Despite this promise, agentic 3D creation remains unreliable: generated results often deviate from user intent and may degrade under iterative refinement.
We argue that these failures stem from two coupled design axes: the \emph{programmatic representation} (what code the agent writes) and the \emph{agent system} (how the agent plans, generates, and improves code).
Early pioneering efforts prompt LLM agents to emit general-purpose modeling programs, including Blender Python scripts, parametric CAD code, or procedural/rule-based programs~\cite{yuan20243d,lu2025ll3m,alrashedy2024generating,du2024blenderllm}.
Although Blender scripts and CAD code are highly expressive, they operate at a low level of abstraction, making it difficult to consistently encode semantic intent; they also tend to be brittle under localized revisions.
In contrast, procedural and rule-based formalisms provide strong parametric control, but often struggle to scale to broad object diversity, stylistic variation, and fine-grained constraints~\cite{raistrick2023infinite,raistrick2024infinigen}.
To improve programmatic representations of 3D assets, recent works introduce Domain-Specific Languages (DSLs) for objects and scenes~\cite{zhang2025scene,zhang2025shapecraft,jones2025aidl}.
While these DSLs improve expressiveness, they still implicitly assume that LLM agents can reliably author and revise valid programs, paying less attention to how the agent system can leverage DSL structure for robust generation, verification, and revision. \looseness=-1

In this work, we address these challenges through the joint design of a programmatic representation for 3D content and a role-specialized multi-agent system.
We observe that LLM agents are effective at decomposing complex 3D assets into hierarchical components and reasoning about spatial and structural relations, yet often struggle to produce precise low-level numeric parameters.
For instance, when generating a chair, an agent can readily infer that the legs should be below the seat and attach at the four corners, but may fail to determine the precise numeric positions of the legs and seat.
Improving the expressiveness of the language alone is insufficient, because even small numeric inaccuracies can yield invalid geometry (e.g., floating legs) or violate design constraints (e.g., missing contact).
To better match agent capabilities, we design a DSL around three complementary aspects:
(1) \emph{expressiveness} to capture diverse shapes and structures;
(2) \emph{composability} to enable modular assembly and hierarchical reuse;
(3) \emph{spatial reasoning} to support precise placement and relational constraints.
Prior work has largely emphasized the first aspect~\cite{zhang2025scene,zhang2025shapecraft}.
Beyond expressiveness, we specifically emphasize the latter two aspects to improve reliability for agent-driven generation and refinement.
With composability, agents can focus on high-level structure while delegating low-level details to reusable components.
With spatial reasoning, agents can invoke relational operators (e.g., ``place A on top of B'') rather than specifying fragile numeric coordinates.
As we design the DSL around agent capabilities, we call it the agent-centric DSL, or \textsc{aDSL}.
\looseness=-1

Built on \textsc{aDSL}, our agent system takes as input a high-level text description or an image specifying the desired 3D content and outputs a program that satisfies the specification.
The system is \emph{training-free} and operates via a self-refinement loop of Plan--Execute--Critic~\cite{yao2022react}.
Instead of learning a specialized policy or relying on task-specific fine-tuning, the agents decompose long-horizon creation into composable, hierarchical subtasks (scene $\rightarrow$ objects $\rightarrow$ parts).
They then generate the program, execute it to obtain concrete geometry and measurable signals, identify violations through automatic checks, and repair the program using feedback from execution logs together with visual and constraint-based evaluations.
The loop continues until all constraints are satisfied or a predefined stopping criterion is met.
Optionally, users can intervene to provide additional guidance or request targeted edits.
Crucially, the \emph{Planner}, \emph{Coder}, and \emph{Critic} communicate through the same representation: the \emph{Planner} emits hierarchical decompositions and checkable relations, the \emph{Coder} realizes them with declarative layout operators, and the \emph{Critic} verifies those same relations after execution.
This shared interface is what we mean by \emph{joint design}; it replaces brittle coordinate-level interaction with a generate--verify--repair loop grounded in the semantics of \textsc{aDSL}. \looseness=-1

We demonstrate that coupling \textsc{aDSL} with an agentic refinement loop improves robustness, controllability, editability, and faithfulness to user intent.
We evaluate our system on text-to-shape and image-to-shape generation tasks, showing gains over recent LLM-based baselines~\cite{siddharth2025blendermcp,du2024blenderllm,lu2025ll3m,zhang2025scene,zhang2025shapecraft} while preserving explicit structure and editability~\cite{Shi2023,zhang2025scene,wu2025direct3ds2}.
These results indicate that our joint design addresses core failure modes in prior agentic 3D creation systems.
We further demonstrate downstream applications, including scene-level generation, articulated object creation, and structured shape editing.
In summary, our key contributions are as follows. \looseness=-1
\begin{itemize}[leftmargin=*,itemsep=2pt]
\item[-] We demonstrate that a shared representation across planning, coding, and critique improves intent faithfulness, controllability, and constraint satisfaction for 3D content creation.
\item[-] We propose an agent-centric DSL for 3D content that combines expressiveness, composability, and spatial reasoning operators.
\item[-] We design a training-free, role-specialized multi-agent system that supports iterative 3D creation and refinement.
\end{itemize}

\section{Related Work} \label{sec:related}


\paragraph{Geometric 3D Representations}
Recent advances in 3D generation have attracted significant attention in academia and industry, driven by large generative models trained on extensive 3D datasets.
A prevailing strategy is to map geometry to compact, learnable representations, including triplanes~\cite{Shue2023,Gupta2023}, 3D Gaussians~\cite{Roessle2024}, sparse voxel grids~\cite{Zheng2023,Ren2024,Xiong2024,Xiang2024,Liu2023,Li2025,He2025,Chen2025}, and latent vector sets~\cite{Zhang2023a,Zhang2024,Hunyuan3d2025}, on which 3D diffusion models operate efficiently.
In parallel, autoregressive models~\cite{Wei2025,Deng2025,Ibing2023,Zhang2022a} cast 3D synthesis as sequence modeling, enabling flexible conditioning and potential scaling behaviors.
While these learning-based pipelines can achieve impressive fidelity at high resolution, they typically require substantial training infrastructure and curated datasets, such as ShapeNet~\cite{Chang2015} or Objaverse~\cite{Deitke2023}.
In contrast, we adopt a training-free, agentic perspective: instead of learning a geometry distribution end-to-end, we leverage the reasoning and tool-use capabilities of LLMs to compose 3D content through programmatic construction. \looseness=-1

\paragraph{Programmatic 3D Representations}
Representing 3D assets as \emph{programs} provides an appealing alternative to raw geometry: programs expose discrete structure, continuous parameters, and explicit compositional hierarchy, making them naturally suited for controllable generation, interpretability, and editing~\cite{jones2020shapeassembly}. \looseness=-1

A straightforward route is to represent 3D content via programs in mature tools, such as Blender Python scripts or parametric CAD code.
With the advent of LLMs, several methods prompt them to synthesize such programs, optionally repairing syntax or runtime errors iteratively~\cite{yuan20243d,lu2025ll3m,alrashedy2024generating,du2024blenderllm}.
These representations are expressive and benefit from mature modeling operators and renderers.
However, they tend to be \emph{too low-level} for specifying high-level semantic intent and are brittle: minor edits can trigger disproportionate geometric changes, and many geometric constraints, such as symmetry and functional relations, remain implicit and difficult to verify without substantial auxiliary tooling or custom checks. \looseness=-1

Procedural modeling has a long history in graphics, from L-systems~\cite{prusinkiewicz2012algorithmic} and shape grammars~\cite{stiny1975pictorial} to urban and architectural generation pipelines \cite{muller2006procedural,krecklau2010generalized,zhang2024cityx}.
By exposing interpretable parameters and enforcing rule-constrained structure, these systems offer strong reliability and controllability, and can efficiently generate diverse variations within a predefined design space.
Their central limitation is the coverage and flexibility: expanding a handcrafted rule set to new object categories, styles, or fine-grained semantic constraints often requires significant expert effort and iterative engineering~\cite{vanegas2012inverse}.
Recent works on large-scale procedural scene generation~\cite{raistrick2023infinite,raistrick2024infinigen} further demonstrate the enduring value of carefully engineered pipelines with procedural modeling components. \looseness=-1

Many works propose domain-specific languages (DSLs) with compositional primitives, higher-level operators, and inductive biases tailored to 3D content.
A prominent line of work represents shapes as compositions of primitives and boolean operations, enabling learning-based inference of programs from data~\cite{Sharma2018,kania2020ucsg,du2018inversecsg,wu2021deepcad,chen2025img2cad}.
Complementary DSLs emphasize part structure and assembly, enabling semantic edits by manipulating structured programs rather than raw meshes~\cite{jones2020shapeassembly}.
GeoCode~\cite{pearl2025geocode} demonstrates that interpretable procedural shape programs can preserve structural validity while supporting high-level edits, and AIDL~\cite{jones2025aidl} introduces a solver-aided hierarchical language for LLM-driven CAD design.
Our work follows these DSL paradigms while also considering agents' capabilities and the needs of 3D creation by exposing explicit constraint semantics that are easy for agents to generate and verify. \looseness=-1

\paragraph{LLM Agents for 3D Creation}
LLM agents have emerged as a general framework for long-horizon problem solving, interleaving planning, tool use, execution, and self-refinement.
Representative paradigms include reasoning--action loops, plan-then-solve decomposition, tree-structured search over intermediate thoughts, and reflection/memory mechanisms~\cite{yao2022react,wang2023plan,yao2023tree,shinn2023reflexion}.
These designs motivate our Plan--Execute--Critic workflow, in which geometric and semantic verification signals provide grounded feedback. \looseness=-1
Recent LLM-based 3D systems cast shape creation as program synthesis, using agents to author Blender scripts~\cite{hu2024scenecraft,lu2025ll3m}, shape programs~\cite{zhang2025shapecraft}, scene-level DSLs~\cite{zhang2025scene}, or parametric CAD programs augmented with visual or verification feedback~\cite{khan2024text2cad,li2024llm4cad,he2025cad,alrashedy2024generating}.
Related efforts further explore indoor and home-scale environments through language parsing, object retrieval, layout optimization, or neural stylization~\cite{ocal2024sceneteller,fu2024anyhome,yang2024holodeck,celen2024idesign,littlefair2025flairgpt,aguinakang2024openuniverse}.
Current work explores procedural scene programs with program-search-based repair~\cite{gumin2025procedural}.
Our method jointly designs the agent workflow and the representation: scene composition, object structure, optional articulation, visual critique, and constraint checking all operate on a shared, verifiable relational program. \looseness=-1

\section{Agentic 3D Creation}
\label{sec:method}

Given a high-level specification of the desired 3D content, such as a text prompt or an image, our goal is to use training-free LLM agents to generate 3D assets as executable programs that are structured, controllable, and interpretable.
We achieve this through the \emph{joint design} of an agent-centric DSL (\textsc{aDSL}) and a multi-agent system that generates, inspects, and repairs DSL programs.
\textsc{aDSL} provides compositional structure and high-level spatial reasoning operators; the \emph{Planner} emits constraints in this vocabulary, the \emph{Coder} realizes them using declarative operators, and the \emph{Critic} re-checks the same relations after execution.
The language therefore serves as a shared interface for generation, evaluation, and correction, rather than a passive output format. \looseness=-1

Our system for agentic 3D creation with \textsc{aDSL} is shown in \Cref{fig:pipeline}.
Given an input specification, the \emph{Planner} derives a structured decomposition of the target object and an explicit set of verifiable constraints.
Conditioned on this plan, the \emph{Coder} synthesizes a program, which is executed and rendered to produce both geometry and visual evidence.
If execution fails, the \emph{Debugger} analyzes runtime errors and proposes targeted repairs to the program.
If execution succeeds, the \emph{Critic} evaluates both the renderings and the program against the \emph{Planner} constraints to produce actionable feedback.
All proposed revisions are fed back to the \emph{Coder}, forming a closed-loop refinement process that improves both executability and adherence to the input specification.
Optionally, user feedback can also be incorporated in the workflow to further guide refinement.
Next, we elaborate on the design of the \textsc{aDSL} and the agent system in \cref{subsec:dsl} and \cref{subsec:workflow}, respectively. \looseness=-1

\subsection{DSL for 3D Assets}
\label{subsec:dsl}

\begin{figure*}[t!]
    \centering
    \includegraphics[width=\textwidth]{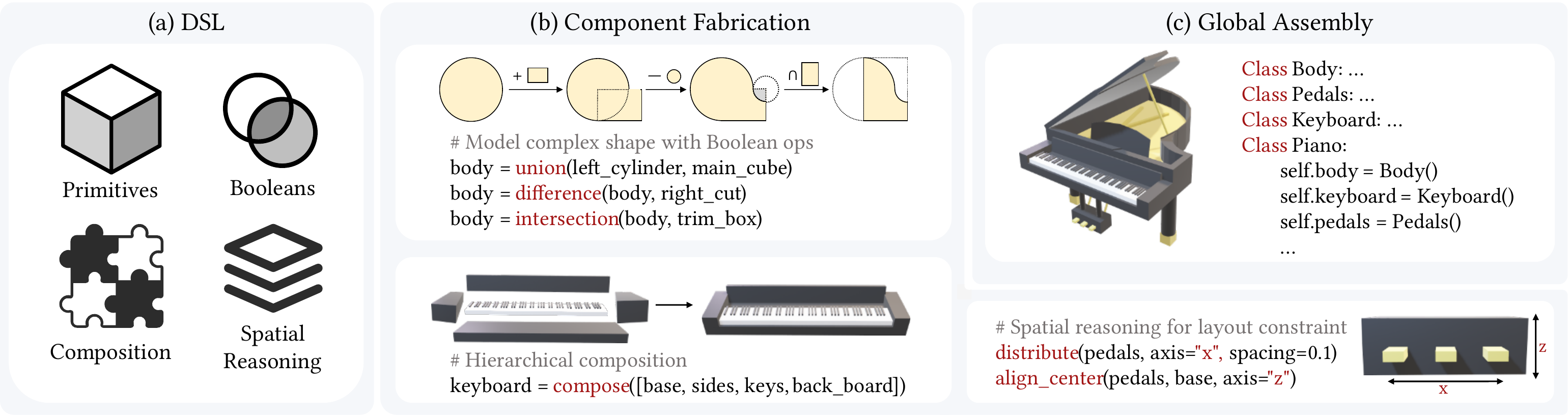}
    \vspace{-6mm}
    \caption{
    Design and use of our \textsc{aDSL} for programmatic 3D shape modeling.
    (a) \textsc{aDSL}: Our framework is built on four core design elements: geometric primitives, boolean operations, hierarchical composition, and spatial reasoning.
    (b) Component Fabrication: Complex shapes are constructed via Constructive Solid Geometry (CSG) and hierarchical composition. Additionally, spatial reasoning operators are applied to declaratively resolve layout constraints and relative positioning.
    (c) Global Assembly: Fabricated parts are organized hierarchically into a semantic object using Python classes, facilitating modular reuse and high-level asset definition.\looseness=-1
    }
    \Description{}
    \label{fig:DSL}

    \vspace{2mm}
    \begin{minipage}[t]{0.48\textwidth}
        \vspace{0pt}
        \centering
        \begin{minipage}[c][0.50\linewidth][c]{\linewidth}
            \centering
            \includegraphics[width=\linewidth]{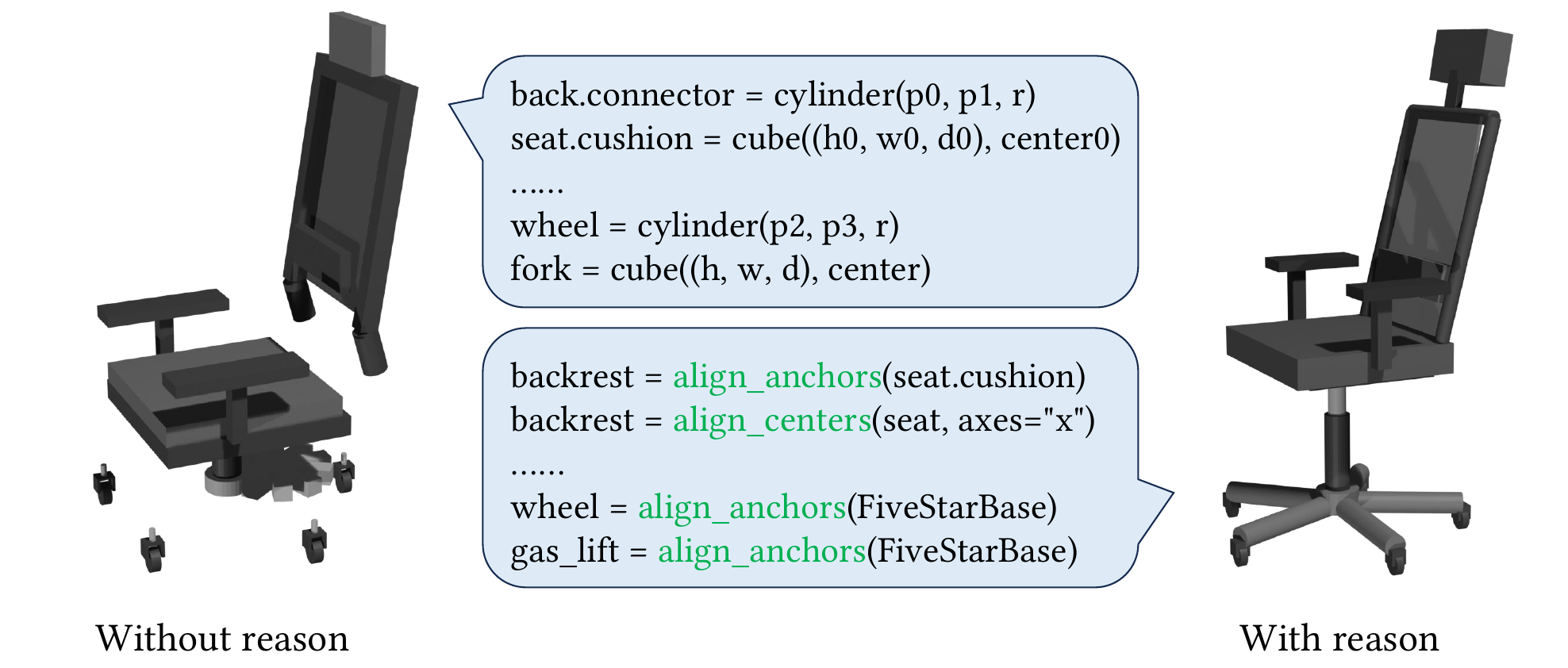}
        \end{minipage}
        \caption{Inter-object spatial constraint handling via spatial reasoning.
        Direct coordinate reasoning can produce disconnected and misaligned components.
        \textsc{aDSL} instead provides declarative operators to align anchors and centers, enabling the agent to encode layout constraints and repair structural violations through interpretable program updates.}
        \Description{}
        \label{fig:reasoning}
    \end{minipage}
    \hfill
    \begin{minipage}[t]{0.48\textwidth}
        \vspace{0pt}
        \centering
        \begin{minipage}[c][0.50\linewidth][c]{\linewidth}
            \centering
            \includegraphics[width=\linewidth]{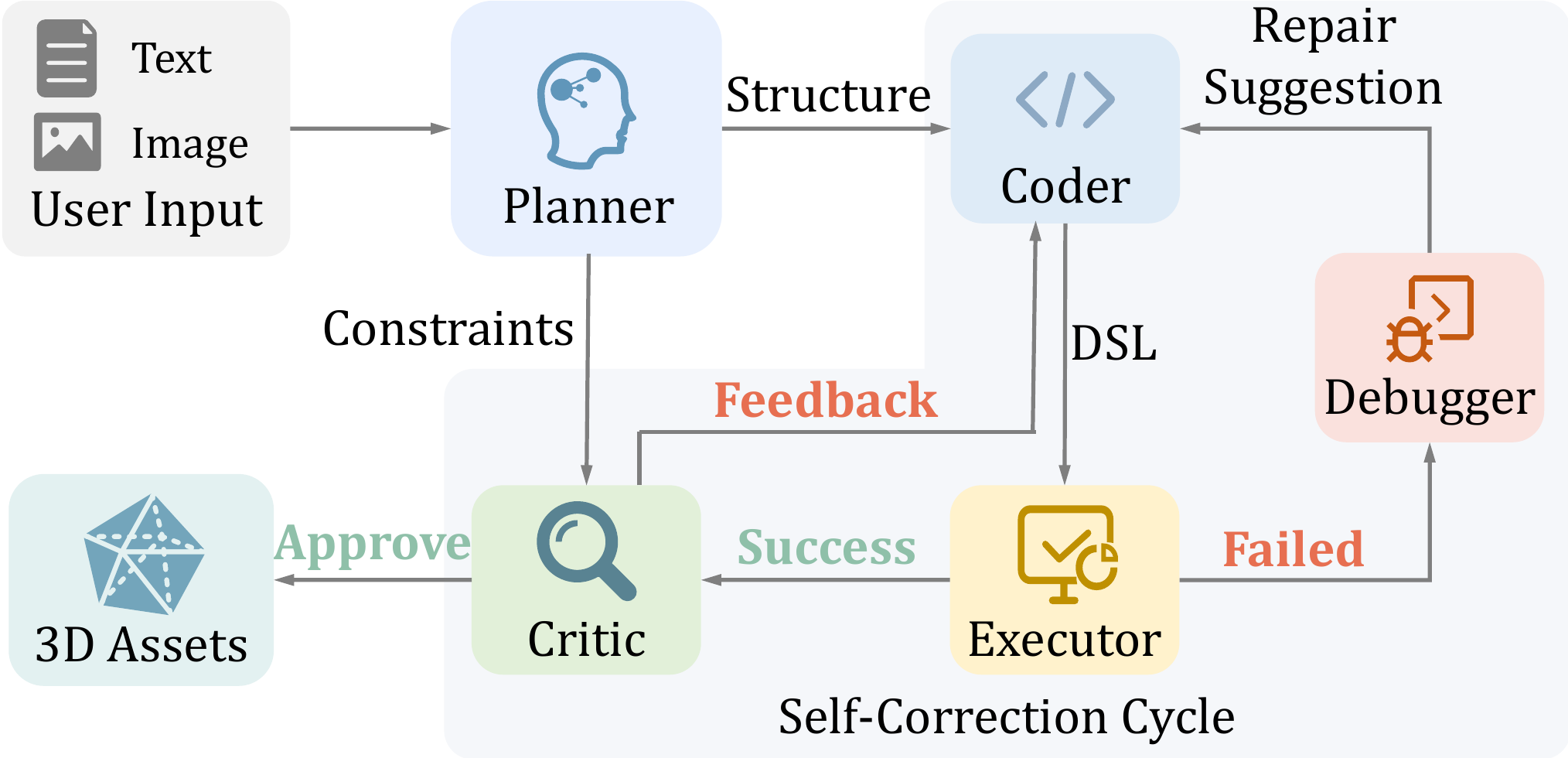}
        \end{minipage}
        \caption{Overview of our agentic 3D creation system.
        Given user input, \emph{Planner} derives structured decompositions and verifiable constraints.
        \emph{Coder} synthesizes a program executed by \emph{Executor} to generate 3D assets.
        Subsequently, \emph{Debugger} resolves execution failures and \emph{Critic} evaluates semantic alignment, providing feedback to refine the generated geometry. \looseness=-1
        }
        \Description{}
        \label{fig:pipeline}
    \end{minipage}
    \vspace{-3mm}
\end{figure*}

\textsc{aDSL} is centered on a hierarchical \texttt{Asset} container with named part attachments, which provides a programmatic substrate for modular decomposition, reusable components, and structured assemblies.
We design \textsc{aDSL} around three key principles: expressiveness, composability, and spatial reasoning.
A representative example is shown in \Cref{fig:DSL}; full syntax and semantics are detailed in the supplementary material.
In particular, expressiveness ensures that a wide variety of shapes can be constructed, while composability and spatial reasoning facilitate LLM-driven synthesis, verification through execution, and iterative refinement. \looseness=-1

\paragraph{Expressiveness}
\textsc{aDSL} first provides core modeling constructs, including parameterized geometric primitives, boolean operators, and geometric transformations.
These basic building blocks support shape creation via constructive solid geometry (CSG)~\cite{foley1996computer} and enable the specification of complex objects via compositional assembly and subtractive refinement.
Specifically, \textsc{aDSL} includes: \looseness=-1
\begin{itemize}[leftmargin=*,itemsep=2pt]
  \item[-] \emph{Primitives}. \textsc{aDSL} includes a compact set of parameterized primitives, such as \texttt{cube}, \texttt{sphere}, and \texttt{cylinder}.
  Each primitive is defined by explicit geometric parameters, with optional appearance attributes, such as color and transparency.
  \item[-] \emph{Boolean operations}. \textsc{aDSL} supports boolean operators, including \texttt{union}, \texttt{intersection}, and \texttt{difference}, enabling part assembly and subtractive carving within a single program by combining intermediate components into progressively refined geometry.
  \item[-] \emph{Transformations}. \textsc{aDSL} supports translation, rotation, scaling, and general affine transforms to control placement and orientation.
\end{itemize}

\paragraph{Composability}
We embed \textsc{aDSL} in Python, reusing its parser, runtime, and familiar abstraction mechanisms, including reusable functions and classes, object hierarchies, and structured control flow such as \texttt{for} loops.
Delegating evaluation to the Python host keeps the DSL compact while preserving deterministic, verifiable execution semantics and providing a natural interface for LLM-based generation, editing, and repair.
Within this embedding, composability is represented by explicit parent-child part attachments: assets are assembled as named hierarchies, enabling component reuse, localized refinement, and incremental construction across levels of detail.
The same hierarchy also supports articulation, since moving parts remain semantically identified within the program.
We attach kinematic relations through parent-part methods such as
{\small\texttt{<parent>\allowbreak.\allowbreak revolute\allowbreak(<child>, ...)}},
where each joint specifies its origin, axis, and motion limits.
Consequently, a single \textsc{aDSL} program defines both the geometry and motion of an asset and can be exported directly as a standardized URDF model, as demonstrated through articulated object creation in \cref{subsec:applications}.
\looseness=-1
\paragraph{Spatial Reasoning}
\textsc{aDSL} augments CSG modeling with a compact layer of \emph{spatial reasoning} primitives for layout- and constraint-driven program synthesis.
For each geometric primitive or composed part, \textsc{aDSL} exposes axis-aligned bounding box (AABB) attributes, including the center, extents, and per-axis minima and maxima, as first-class relational queries.
These AABB queries are used only to describe relations and layout constraints.
The underlying geometry is still represented by primitives, CSG operations, and geometric transformations, allowing the modeling of non-axis-aligned and more complex shapes.
Built on these queries, \textsc{aDSL} provides \emph{declarative} layout operators, such as placement, center alignment, and distribution, that express common spatial relations as single program statements rather than brittle, low-level numeric choices.
These operators naturally support an agentic generate--verify--repair loop.
During generation, the agent translates high-level requirements into explicit spatial constraints and selectively applies the appropriate operators to satisfy them.
After execution, the resulting geometry can be checked through both renderings and the program state.
When violations are detected, the program is repaired by adjusting operator arguments (offsets, axes, ordering, and distribution parameters) or inserting additional layout steps, and the process repeats until all constraints are satisfied.
By routing synthesis through these declarative operators, \textsc{aDSL} reduces reliance on fragile numeric values and improves reliability in our experiments.
An example of this advantage is shown in \cref{fig:reasoning}.
\looseness=-1

\paragraph{Remarks}
LL3M~\cite{lu2025ll3m} builds an agent system that generates low-level Blender Python scripts for 3D modeling, while Scene Language~\cite{zhang2025scene} studies DSL design for 3D scenes; neither directly addresses spatial reasoning as a first-class substrate for agentic editing.
In contrast, \textsc{aDSL} combines a Python hierarchy interface and CSG-style expressiveness with declarative spatial operators, enabling constraint-driven synthesis and post-execution verification.
AIDL~\cite{jones2025aidl} also recognizes the spatial reasoning limitations of LLMs and augments them with a geometric constraint solver, but its focus is primarily on 2D CAD sketches.
For complex 3D content creation, global solvers can be brittle under \emph{local refinements} and often provide limited semantic feedback.
\textsc{aDSL} instead exposes spatial intent through interpretable program primitives, giving the agent actionable feedback for systematic checking and repair throughout the generation and editing loop. \looseness=-1

\subsection{Agent System}
\label{subsec:workflow}

In this section, we present our agent system that orchestrates role-specialized agents to generate, verify, and refine 3D assets as \textsc{aDSL} programs.
The system consists of four stages: planning, coding and execution, critique, and memory/context management.
Each stage is handled by dedicated agents, as illustrated in \cref{fig:pipeline}. \looseness=-1

\paragraph{Planning Stage}
The workflow starts with a planning stage that translates the user request into an explicit, checkable specification.
The \emph{Planner} parses the input and outputs a modeling specification with three components: \looseness=-1
\begin{itemize}[leftmargin=*,itemsep=2pt]
\item[-] \emph{Component decomposition:} a hierarchical decomposition of the target asset, with natural-language descriptions that guide geometric construction and align with \textsc{aDSL}'s compositional structure;
\item[-] \emph{Spatial relations:} constraints on connectivity, alignment, and relative placement among components, expressed in a form that can be partially mapped to \textsc{aDSL}'s spatial reasoning primitives;
\item[-] \emph{Critic checklist:} a set of precise, verifiable criteria derived from the user requirements, including component existence, counts, support/contact relations, and alignment constraints.
\end{itemize}
We forward the component decomposition and spatial relations to the \emph{Coder} to ground implementation in a stable architectural plan, and provide the checklist to the \emph{Critic} for systematic verification and targeted revision. \looseness=-1

\paragraph{Coding and Execution Stage}
Conditioned on the \emph{Planner} outputs, the \emph{Coder} synthesizes an \textsc{aDSL} program by instantiating primitives, composing them into a named part hierarchy, and applying transformations and layout operators to satisfy the specified constraints.
The \emph{Executor} runs the program to generate geometry and export it to a mesh representation.
Upon successful execution, the renderer captures visual evidence for downstream assessment by producing multi-view snapshots that minimize self-occlusion and improve coverage of local geometric details.
If execution fails (e.g., due to invalid parameters, missing definitions, or malformed operator usage), the \emph{Debugger} analyzes the error signals and proposes targeted patches.
These repairs are returned to the \emph{Coder} for revision, grounding refinement in observable program behavior and ensuring executability before critique. \looseness=-1

\paragraph{Critique Stage}
Upon successful execution, the workflow enters a refinement phase that couples visual inspection with program-level verification.
First, an \emph{Image Critic} compares multi-view renderings to the \emph{Planner}'s checklist, identifying perceptual and structural discrepancies (e.g., missing components or incorrect proportions), while ignoring minor rendering artifacts.
These visual observations are forwarded to a \emph{Code Critic}, which serves as the final adjudicator.
The \emph{Code Critic} cross-references the visual feedback with the underlying \textsc{aDSL} program, using the same code structure to verify the validity of the reported issues.
This verification ensures that the self-correction loop is driven by programming faults rather than hallucinations caused by occlusion or perspective ambiguity. \looseness=-1

\paragraph{Memory and Context Management}
To prevent context overflow while ensuring long-term stability during the iterative refinement process, we use a \emph{selective memory} mechanism that separates persistent constraints from transient working state.
User input and the \emph{Planner}'s output form \emph{persistent memory}, preserved across all agents to ensure the original goal is never lost.
All other intermediate results are treated as \emph{transient memory} and pruned by specific rules.
The \emph{Coder} maintains a sliding window that retains only fixed requirements and the most recent code synthesis, while discarding stale code and debug logs to keep the workspace clean.
The \emph{Critic} enforces a strict reset policy for ``data-heavy'' content: specifically, multi-view renderings are removed from history after each round, while the textual record of prior feedback is preserved.
This design prevents contamination by obsolete visuals and preserves cross-round continuity in the feedback stream. \looseness=-1

\section{Results} \label{sec:exp}

We first describe the experimental setup, then report text-to-shape and image-to-shape results, followed by ablations and downstream applications in articulated object generation, shape editing, high-fidelity generation, scene-level modeling, and user interaction. \looseness=-1


\paragraph{Baselines}
We compare against state-of-the-art baselines from three generation paradigms:
\begin{itemize}[leftmargin=*,itemsep=2pt]
\item[-] \emph{Code generation}: BlenderMCP~\cite{siddharth2025blendermcp}, BlenderLLM~\cite{du2024blenderllm}, LL3M~\cite{lu2025ll3m}, Scene Language~\cite{zhang2025scene}, and ShapeCraft~\cite{zhang2025shapecraft}, which generate code to produce meshes. \looseness=-1
\item[-] \emph{Field generation}: MVDream~\cite{Shi2023}, LN3Diff~\cite{lan2024ln3diff}, Trellis~\cite{Xiang2024}, and Direct3D-s2~\cite{wu2025direct3ds2}, which generate implicit fields that are later converted to meshes. \looseness=-1
\item[-] \emph{Mesh generation}: Llama-Mesh~\cite{Wang2024a}, which directly outputs triangle meshes. \looseness=-1
\end{itemize}
Field and mesh generation methods rely on large-scale 3D training data, which is fundamentally different from our \emph{training-free} code generation approach; we therefore treat them as reference comparisons that contextualize our method within the broader 3D generation landscape.
LL3M is evaluated qualitatively and BlenderMCP is evaluated quantitatively on a subset due to limited API access quotas.
Implementation details and metric definitions are provided in supplmentary materials.
\looseness=-1

\subsection{Text-to-Shape Generation}
\label{subsec:text2shape}

\paragraph{Datasets}
We construct a benchmark of $100$ randomly sampled text-conditioned instances: $60$ from ShapeNet~\cite{Chang2015}, $20$ from ABO~\cite{collins2022abo}, and $20$ from Objaverse~\cite{Deitke2023}.
Each instance is paired with two prompt templates from CAP3D~\cite{Luo2023} and MARVEL~\cite{sinha2025marvel}, yielding $200$ evaluation prompts that cover complementary linguistic descriptions.
To ensure a controlled comparison, all text-to-shape methods use the original prompts without additional prompt engineering.
For open-ended user inputs, our framework can optionally prepend an agent that converts concise user requests into structured modeling specifications. \looseness=-1

\paragraph{Quantitative Results}
\cref{tab:text2shape} summarizes the quantitative performance across all methods.
Our method achieves the best performance among code-generation approaches, outperforming recent baselines such as Scene Language~\cite{zhang2025scene} and ShapeCraft~\cite{zhang2025shapecraft} on both CLIP and VQA metrics, while maintaining a 100\% execution success rate.
This gain comes from using \textsc{aDSL} as a shared representation for generation and verification: the \emph{Planner} specifies checkable spatial relations and hierarchical structure, the \emph{Coder} realizes them with declarative operators, and the \emph{Critic} verifies the same relations after execution.
Compared with raw Blender scripts, which require LLMs to manipulate low-level API calls and fragile coordinates, \textsc{aDSL} preserves structure, editability, and user-level intent throughout synthesis, enabling violations to be detected and repaired easily.
\looseness=-1

\begin{figure*}[p]
    \centering
    \includegraphics[width=\textwidth]{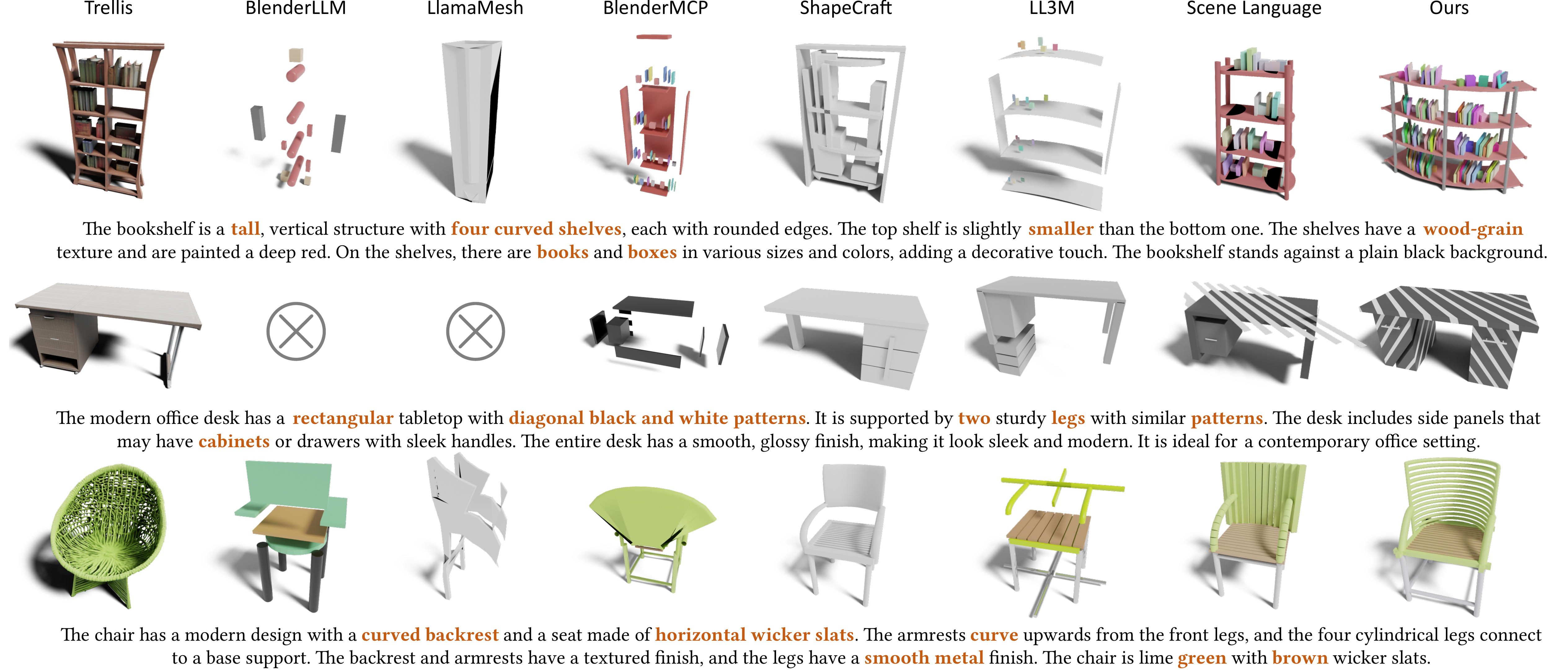}
    \caption{Qualitative comparison on text-to-shape generation. Cross marks indicate mesh generation failures.}
    \label{fig:text2shape}
\end{figure*}

\begin{figure*}[p]
    \centering
    \includegraphics[width=\textwidth]{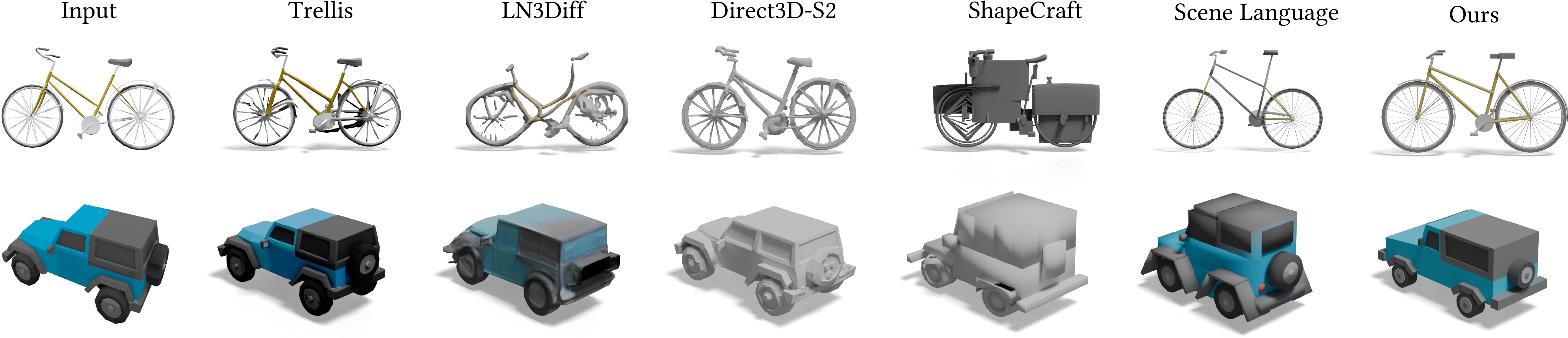}
    \caption{Qualitative comparison of image-to-shape generation.}
    \label{fig:image2shape}
\end{figure*}

\begin{figure*}[p]
    \centering
    \includegraphics[width=\textwidth]{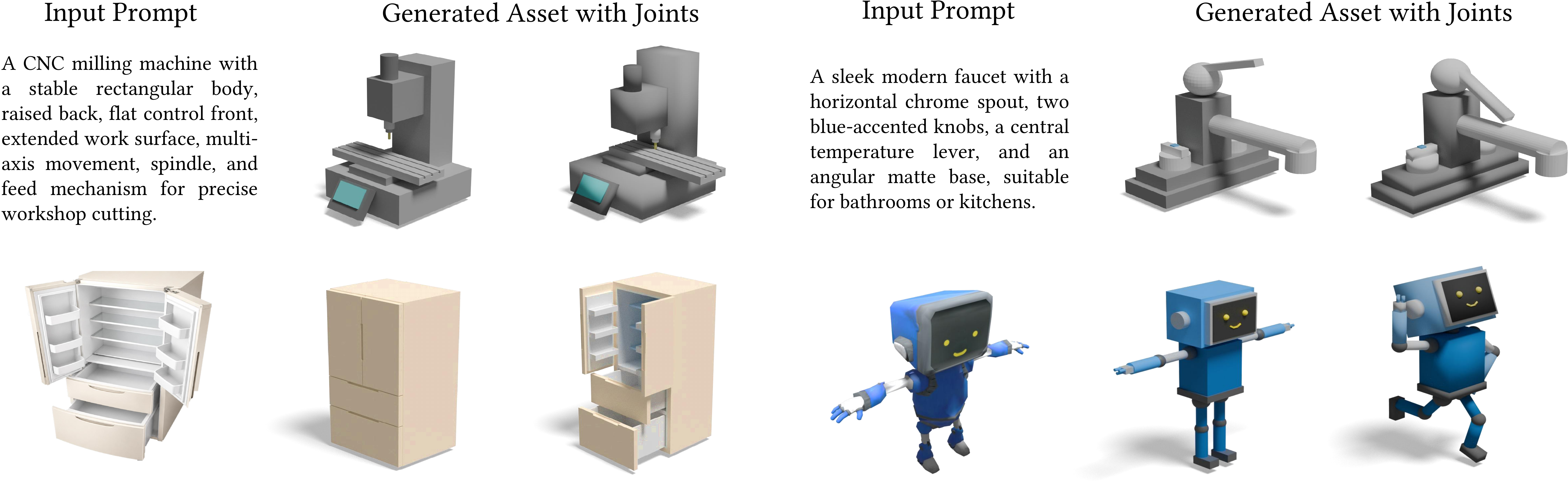}
    \caption{Articulated object generation results. Given text or image prompts, our system synthesizes structured assets together with joint-enabled part hierarchies, covering diverse articulated objects including industrial tools, appliances, and characters.\looseness=-1 }
    \label{fig:articulation}
\end{figure*}

\begin{figure*}[t!]
    \centering
    \makeatletter\def\@captype{table}\makeatother
    \begin{minipage}[t]{0.48\textwidth}
        \vspace{0pt}

    \centering
    \tablestyle{1.4pt}{1.1}
    \captionof{table}{Quantitative results on text-to-shape generation.}
    \vspace{-3mm}
    \label{tab:text2shape}
    \begin{tabular}{l ccc ccc ccc}
        \toprule
        \multirow{2}{*}{Method} &
            \multicolumn{3}{c}{ShapeNet} &
            \multicolumn{3}{c}{ABO} &
            \multicolumn{3}{c}{Objaverse} \\
        \cmidrule(lr){2-4}\cmidrule(lr){5-7}\cmidrule(lr){8-10}
         & CLIP$\uparrow$ & VQA$\uparrow$ & Succ.$\uparrow$ & CLIP$\uparrow$ & VQA$\uparrow$ & Succ.$\uparrow$ & CLIP $\uparrow$ & VQA $\uparrow$ & Succ. $\uparrow$ \\
        \midrule
        Llama-Mesh                        & 20.29 & 45.15 & 0.85 & 22.63 & 53.55 & 0.90 & 17.29 & 38.74 & 0.80 \\
        \midrule
        Trellis          & 28.29 & 68.34 & 1.00 & 29.50 & 66.94 & 1.00 & 27.67 & 68.06 & 1.00 \\
        LN3Diff                                  & 22.96 & 48.10 & 1.00 & 24.04 & 55.46 & 0.95 & 21.68 & 52.52 & 1.00 \\
        MVDream                                & 23.36 & 60.57 & 1.00 & 22.78 & 61.96 & 0.98 & 22.16 & 59.08 & 0.88 \\
        \midrule
        \scriptsize{Scene Language}  & 28.35 & 59.13 & 0.97 & 29.18 & 65.42 & 0.98 & 26.77 & 62.45 & 0.95 \\
        ShapeCraft                               & 27.70 & 57.26 & 1.00 & 29.08 & 62.85 & 1.00 & 24.60 & 53.94 & 1.00 \\
        BlenderLLM                               & 24.01 & 52.81 & 0.91 & 25.80 & 61.03 & 0.95 & 21.69 & 48.72 & 0.88 \\
        BlenderMCP                               & --    & --    & --   & 28.97 & 64.76 & 1.00 & 28.44 & 66.22 & 1.00 \\
        \textsc{aDSL} (Ours)                     & \textbf{29.63} & \textbf{65.34} & \textbf{1.00} & \textbf{30.39} & \textbf{68.10} & \textbf{1.00} & \textbf{29.07} & \textbf{69.37} & \textbf{1.00} \\
        \bottomrule
    \end{tabular}
    \vspace{-2mm}

        \vspace{6mm}
            \centering
    \tablestyle{10pt}{1.1}
    \captionof{table}{Efficiency statistics for text-conditioned shape generation under our refinement stopping protocol.}
    \label{tab:efficiency}
    \vspace{-3mm}
    \begin{tabular}{
        l
        S[table-format=3.2]
        S[table-format=3.2]
        S[table-format=3.2]
    }
        \toprule
        Metric & \multicolumn{1}{c}{ShapeNet} & \multicolumn{1}{c}{ABO} & \multicolumn{1}{c}{Objaverse} \\
        \midrule
        Time / Round (s)          & 195.12 & 187.50 & 190.26 \\
        Input Tokens / Round (k)  & 30.87  & 30.92  & 32.57  \\
        Output Tokens / Round (k) & 2.75   & 2.59   & 2.99   \\
        Average Rounds            & 4.25   & 4.50   & 5.23   \\
        \bottomrule
    \end{tabular}
    \vspace{-4mm}

    \end{minipage}
    \hfill
    \begin{minipage}[t]{0.48\textwidth}
        \vspace{0pt}
        \tablestyle{11pt}{1.35}
\captionof{table}{Quantitative results for image-to-shape generation on Toys4K.}
\vspace{-3mm}
\label{tab:image2shape}
\begin{tabular}{llccc}
    \toprule
    Method          & Type                 & CLIP $\uparrow$ & FID$_\text{incp}$ $\downarrow$ & Succ. $\uparrow$ \\
    \midrule
    Trellis         & \multirow{2}{*}{Field} & 84.88 & 108.20 & 1.00 \\
    Direct3D-S2     &                        & 82.13 & 148.62 & 1.00 \\
    \midrule
    Scene Language  & \multirow{4}{*}{Code}  & 78.68 & 206.21 & 0.93 \\
    ShapeCraft      &                        & 79.34 & 187.62 & 1.00 \\
    BlenderMCP      &                        & 83.28 & 214.87 & 1.00 \\
    \textsc{aDSL} (Ours)             &                        & \textbf{84.42} & \textbf{184.71} & 1.00 \\
    \bottomrule
\end{tabular}
\vspace{-4mm}

        \vspace{6mm}
        \tablestyle{7pt}{1.35}
\captionof{table}{Ablation of the relational program interface and iterative agentic workflow on text-conditioned shape generation.}
\label{tab:ablation}
\vspace{-3mm}
\begin{tabular}{lcccc}
    \toprule
    Method                          & CLIP $\uparrow$ & VQA $\uparrow$ & Succ. $\uparrow$ & Rounds $\downarrow$ \\
    \midrule
    w/o Spatial Utils              & 29.38 & 63.75 & 1.00 & 4.67 \\
    Blender Script                 & 28.11 & 62.99 & 1.00 & 6.08 \\
    w/o \emph{Planner}             & 29.52 & 64.12 & 1.00 & 5.58 \\
    w/o Refinement                 & 29.00 & 61.53 & 0.98 & - \\
    w/o Spatial Utils \& Refinement & 28.20 & 59.25 & 0.97 & - \\
    \midrule
    \textsc{aDSL} (Ours)                            & \textbf{29.63} & \textbf{65.34} & 1.00 & \textbf{4.25} \\
    \bottomrule
\end{tabular}
\vspace{-5mm}

    \end{minipage}
\end{figure*}

\paragraph{Qualitative Results}

\cref{fig:text2shape} compares our method with representative baselines.
Our method produces shapes that better preserve the input semantics while maintaining coherent structure and valid geometry.
Field-based methods such as Trellis can generate visually rich results, but often miss fine-grained constraints, e.g., the ``4 curved shelves'' of the bookshelf or the ``diagonal black and white patterns'' on the desk.
Compared with code-generation baselines, our explicit relational structure and iterative repair loop reduce layout errors such as floating components and misaligned parts. \looseness=-1

\paragraph{Efficiency}
\cref{tab:efficiency} reports the average running time and token usage for text-to-shape generation across ShapeNet, ABO, and Objaverse.
The system takes approximately 190s per round, and on average requires 4.7 rounds to converge to a valid solution, resulting in a total time of around 889s per object.
On average, more than 95\% of the runtime is spent on LLM responses, with the remaining overhead dominated by rendering and execution.
For reference, ShapeCraft~\cite{zhang2025shapecraft} reports an average runtime of 700s per object, while LL3M~\cite{lu2025ll3m} reports a generation time of $\approx 10$ minutes per object, placing our full pipeline in a comparable wall-clock range.
In practical interactive use, however, the effective latency can be substantially lower.
The first request for a complex asset is typically the most expensive, since the agent must construct the program structure from scratch.
After the initial request establishes the program structure, follow-up requests can reuse prior code and modeling decisions.
\cref{fig:continuous-chat} illustrates this behavior: the initial motorcycle requires five refinement rounds and 845s, whereas a follow-up cyber-punk variant reuses the existing context, completes in one round, and takes only 164s. \looseness=-1

\paragraph{Human Evaluation}
To complement automatic metrics, we also conduct a pairwise user study against Scene Language~\cite{zhang2025scene} on 20 cases: 15 text-to-shape and 5 image-to-shape instances (\cref{subsec:image2shape}).
We recruit $38$ participants and evaluate two criteria: prompt alignment, covering semantics, relations, and attributes; and geometric/visual quality, covering plausibility, completeness, and artifacts.
For each case, participants see the input prompt and randomly ordered renderings from both methods, then select the result that better satisfies each criterion.
Overall, participants prefer our results in $85.39\%$ for prompt alignment and $86.84\%$ for geometric/visual quality, confirming that the improvements are perceptually salient and not merely artifacts of automatic metrics. \looseness=-1

\suppressfloats[t]
\subsection{Image-to-Shape Generation}
\label{subsec:image2shape}

\paragraph{Datasets.}
We randomly sample $30$ instances from Toys4K~\cite{stojanov2021using}, which contains diverse rigid object categories and is used by recent image-conditioned baselines such as Trellis.
Each object is rendered from a random viewpoint as the input condition, and all methods are evaluated without additional text descriptions or prompt expansion, ensuring a pure image-to-shape protocol. \looseness=-1

\paragraph{Quantitative Results}
\cref{tab:image2shape} summarizes the quantitative performance across all methods.
Our method achieves the best performance among code-generation approaches, outperforming Scene Language~\cite{zhang2025scene} and ShapeCraft~\cite{zhang2025shapecraft} on both CLIP and FID.
As in text-to-shape generation, the gains come from coupling explicit relational structure with iterative visual repair, which improves image alignment while preserving controllable and editable outputs. \looseness=-1

\paragraph{Qualitative Results}
\Cref{fig:image2shape} compares our method with representative baselines on image-to-shape generation.
Compared with other code-generation baselines, our method produces structurally cleaner outputs without floating parts or interpenetrating components, and remains more consistent with the reference image.
Field-based methods still recover smoother surfaces and richer appearance cues, but they struggle with structured details such as bicycle spokes.
This limitation is complementary to the strengths of program synthesis, which emphasizes explicit structure, editability, and functional correctness.
\Cref{subsec:applications} shows how the two paradigms can be combined to obtain both high fidelity and controllability. \looseness=-1

\suppressfloats[t]
\subsection{Ablation Studies}
\label{subsec:ablation}

We ablate the main components of our representation and agentic workflow on the ShapeNet subset of the text-to-shape benchmark, using $120$ evaluation prompts.
\Cref{tab:ablation} reports quantitative performance and the average number of refinement rounds required to reach a valid solution. \looseness=-1

\paragraph{3D Modeling Language.}
We evaluate the impact of our proposed DSL by
(i) removing the spatial reasoning utilities and declarative layout operators, thereby forcing the model to rely on manual coordinate arithmetic; and
(ii) replacing it with raw Blender Python scripting while preserving the agentic framework, which tests whether the gains arise from the DSL design.
As shown in \cref{tab:ablation}, Blender scripting significantly increases the average number of self-correction rounds from 4.25 to 6.08.
This indicates that while standard Blender scripting is expressive, it takes more effort to converge to a valid and semantically accurate solution.
Similarly, removing spatial utilities increases the average number of iterations to 4.67 and causes a notable drop in VQA score (65.34 $\rightarrow$ 63.75), suggesting that explicit layout operators are helpful for satisfying complex spatial constraints. \looseness=-1

\paragraph{Agentic Workflow.}
We assess the effectiveness of our workflow by
(i) removing the planning stage, where the \emph{Coder} generates programs directly without structured decomposition and the \emph{Critic} lacks a consistent checklist for verification; and
(ii) disabling the self-correction loop, restricting the system to single-pass execution.
The results in \cref{tab:ablation} demonstrate that the planning stage is crucial for efficiency:
removing it increases the average number of refinement rounds from 4.25 to 5.58.
Most critically, disabling the self-correction loop results in a low VQA score (61.53) and a drop in execution success rate to 0.98, highlighting that iterative verification is indispensable for ensuring both the semantic fidelity and structural validity of the generated assets.
\looseness=-1

\paragraph{Joint Effects.}
The combined ablation further shows that the improvement comes from \emph{joint design} rather than either component alone.
Removing refinement preserves declarative spatial operators but prevents the system from repairing missed constraints, while removing spatial utilities keeps iterative correction but forces revisions into brittle low-level coordinate edits.
When both are removed, performance drops further to 28.20 CLIP, 59.25 VQA, and 0.97 success rate, which is worse than either individual ablation.
This indicates that spatial operators and iterative refinement are complementary: the DSL exposes relations in a form that is easy to verify and revise, while the refinement loop turns that structure into reliable error correction.
Taken together, these results support our central claim that robustness arises from coupling an LLM-friendly representation with an agentic repair process. \looseness=-1

\begin{figure}[t!]
    \centering
    \includegraphics[width=\linewidth]{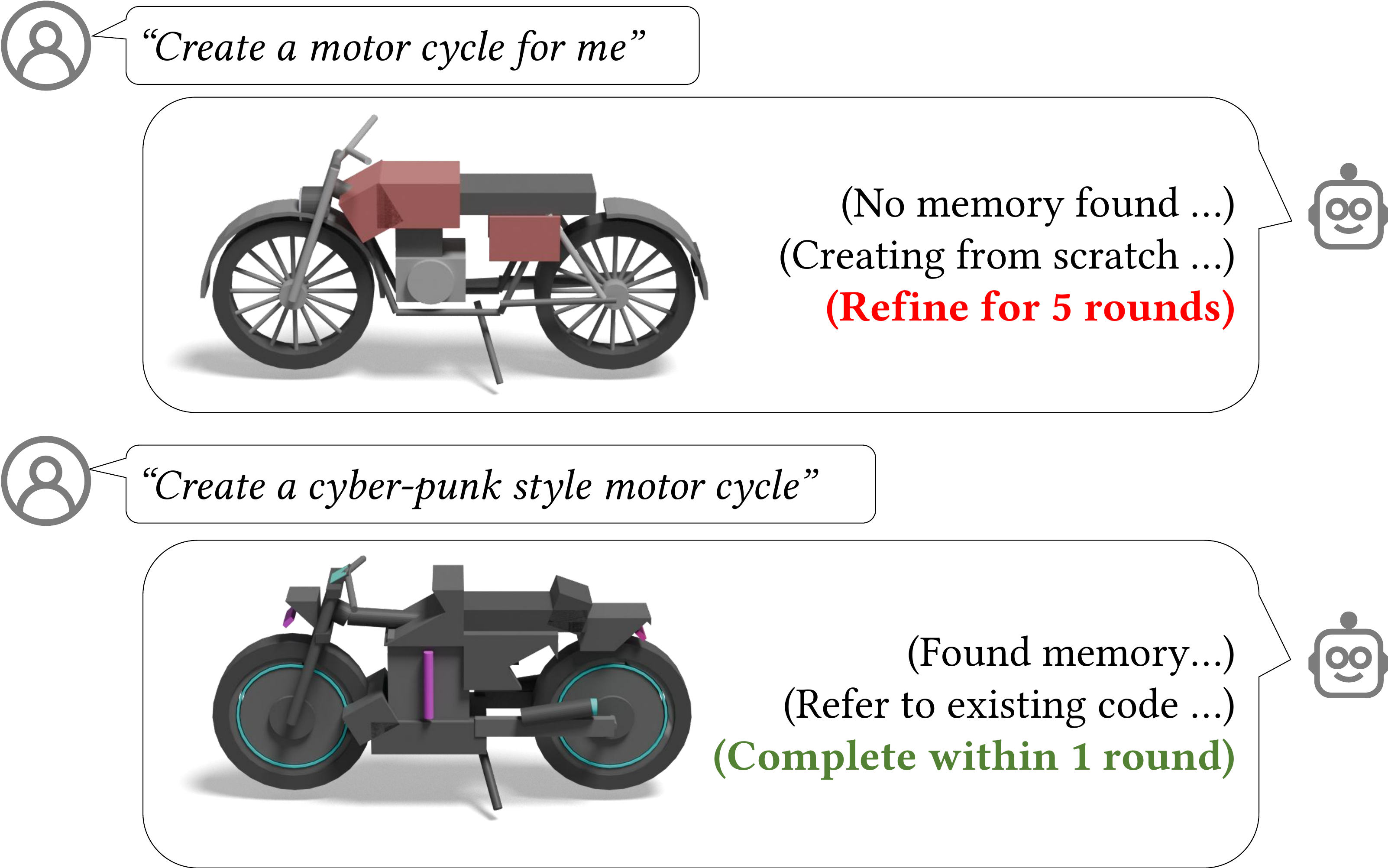}
    \caption{Efficiency gain from continuous interaction with memory reuse. The first user request is generated from scratch and requires five refinement rounds, while the follow-up request reuses the prior solution from memory and completes the generation \emph{in one round}.}
    \label{fig:continuous-chat}
    \vspace{2mm}
    \includegraphics[width=\linewidth]{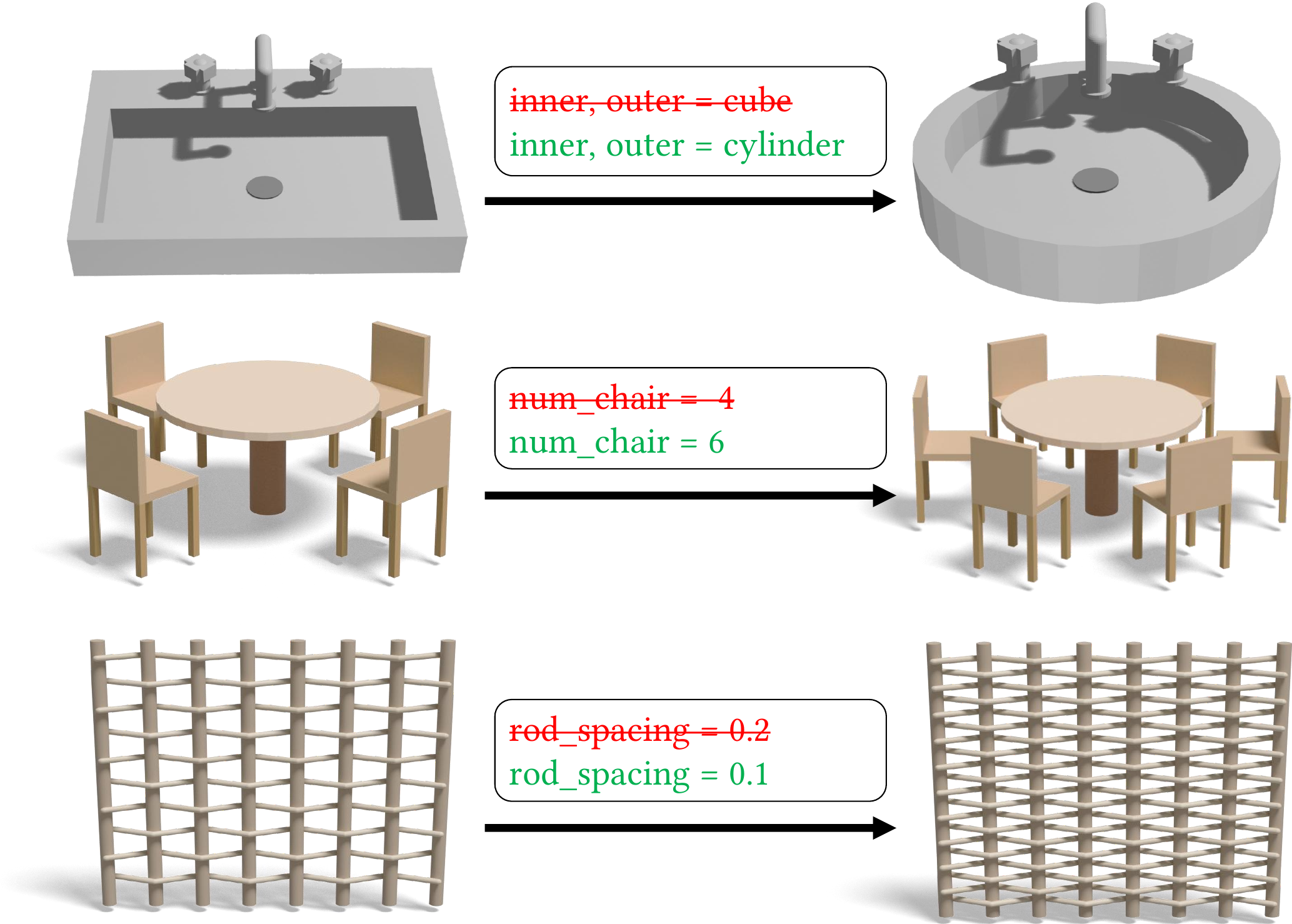}
    \caption{Shape editing via localized program rewrites.}
    \label{fig:editing}
\end{figure}

\begin{figure}[t!]
    \centering
    \includegraphics[width=.95\linewidth]{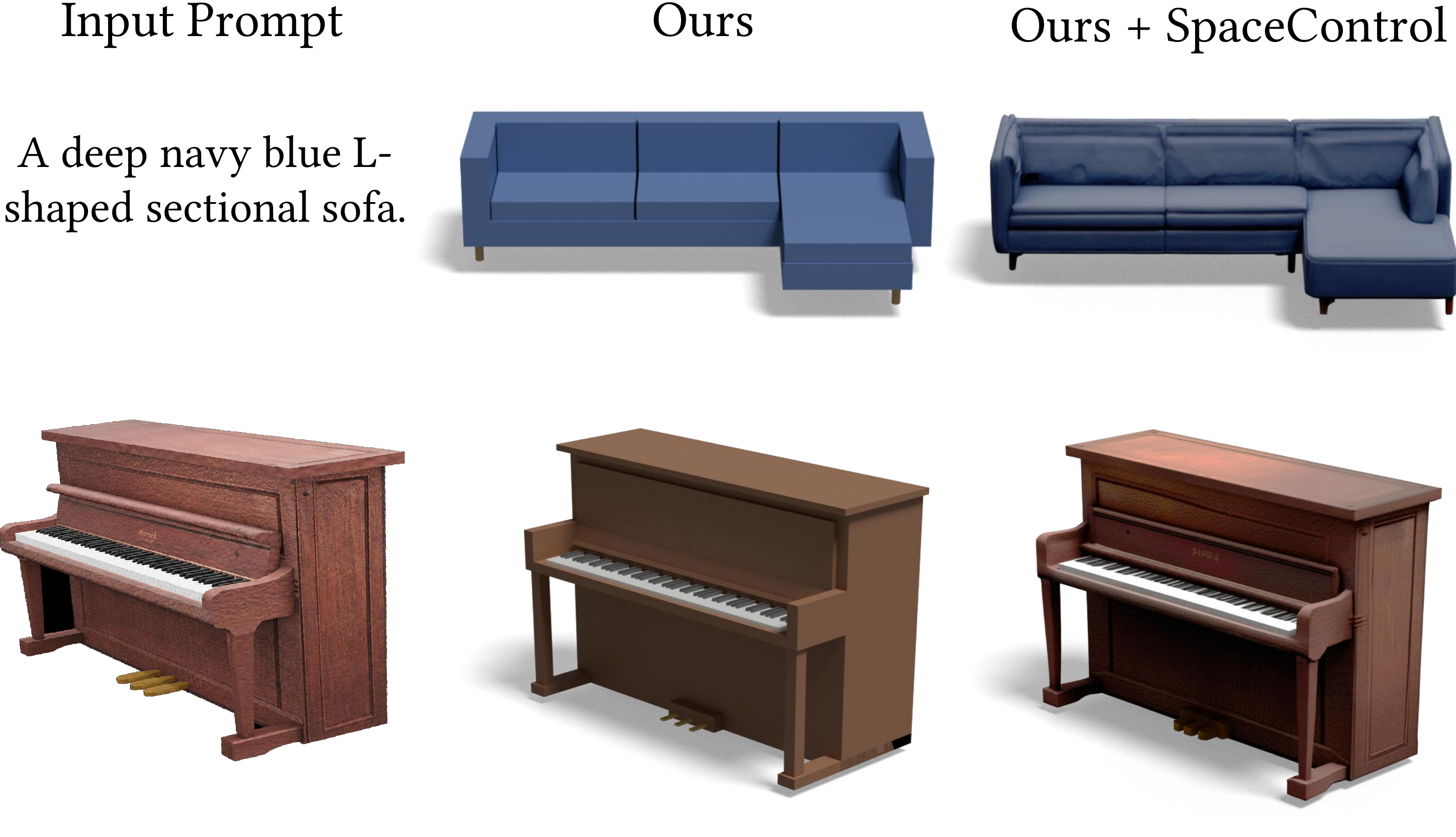}
    \vspace{-15pt}
    \caption{High-fidelity shape generation results via external model conditioning. The \textsc{aDSL} program (``Ours'') serves as a geometric condition to guide the external generator via the SpaceControl protocol (``Ours + SpaceControl'').\looseness=-1}
    \label{fig:high-fidelity}
    \includegraphics[width=.95\linewidth]{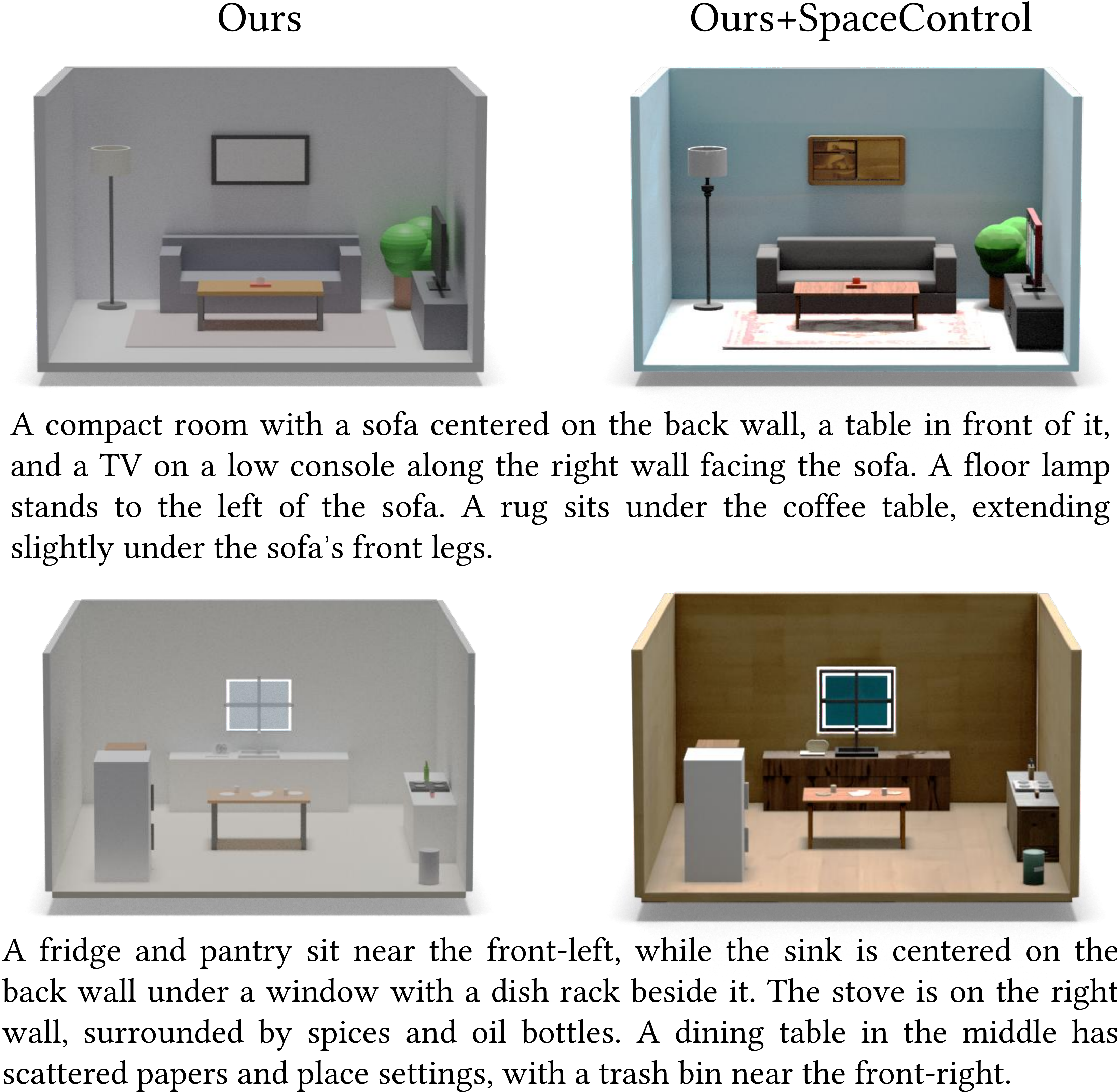}
    \caption{High-fidelity text-to-scene generation results. Our hierarchical \textsc{aDSL} separates object structure from scene layout, enabling objects to be extracted, refined with a pre-trained generator, and recomposed into a high-fidelity scene under the original spatial constraints.}
    \label{fig:scene-composition}
    \includegraphics[width=.95\linewidth]{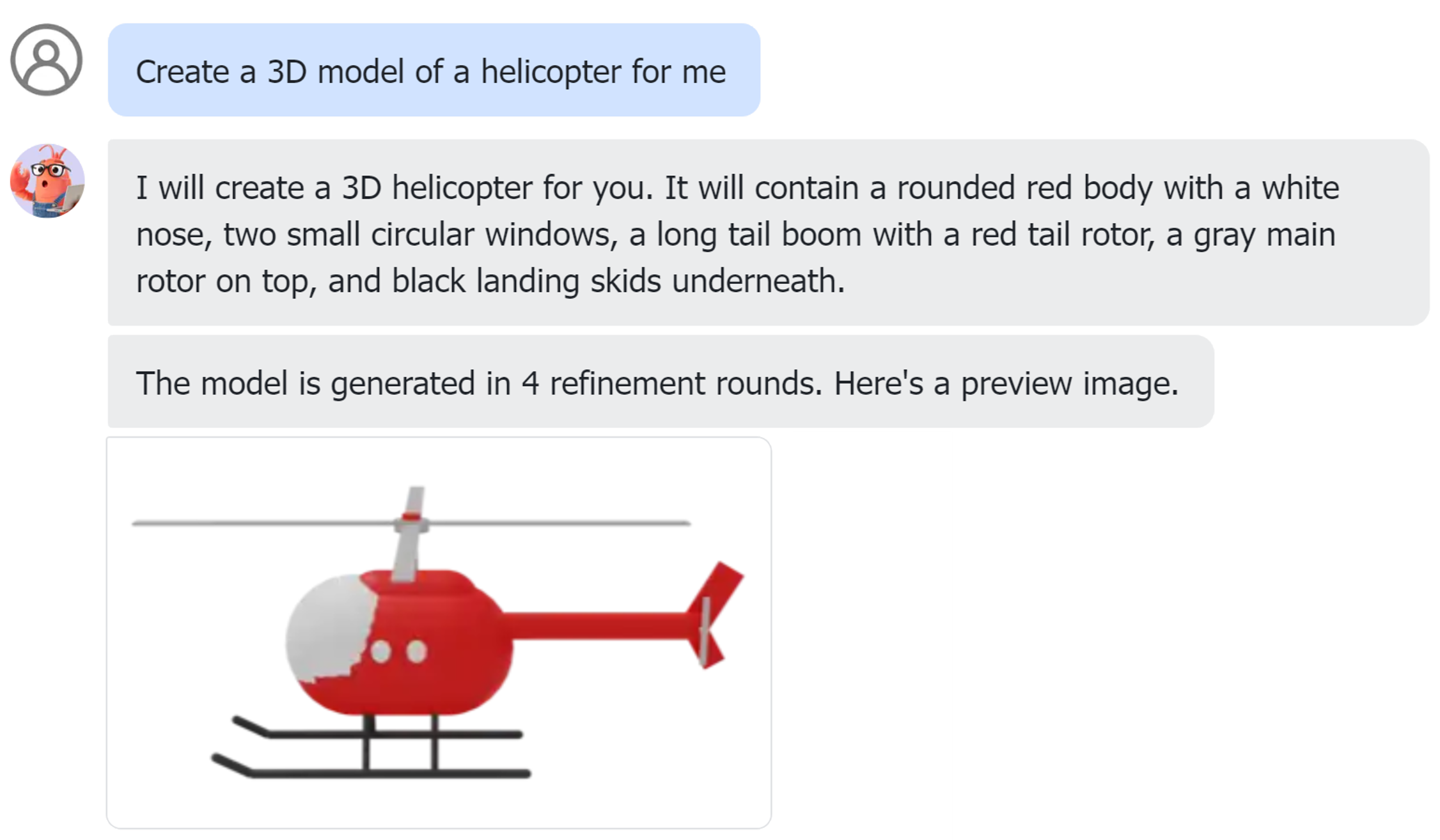}
    \caption{Interactive integration with OpenClaw.}
    \label{fig:openclaw}
\end{figure}

\subsection{Applications}
\label{subsec:applications}


\paragraph{Articulated Shape Generation}
\textsc{aDSL} can encode geometry and kinematics within a unified program, enabling the synthesis of articulated objects with explicit part hierarchies and joint definitions.
\Cref{fig:articulation} illustrates this capability across diverse articulated structures, such as sliding components, rotating handles, hinged doors, and articulated limbs.
These examples demonstrate that our \textsc{aDSL} provides a robust foundation for modeling both the visual geometry and the underlying mechanical function of complex 3D objects.
More results are provided in the supplementary video.
\looseness=-1

\paragraph{Shape Editing}
We formulate shape editing as localized program rewriting rather than regeneration.
Given an instruction and an existing DSL program, the agent identifies the relevant parameters and updates only affected statements.
\Cref{fig:editing} shows controlled edits to primitive type, object count, and spacing, where intended components change while unaffected geometry and connectivity are preserved.
Each visual change therefore corresponds to an explicit, interpretable code revision. \looseness=-1

\paragraph{High-Fidelity Shape Generation}
We couple the structured DSL scaffold with an external 3D generator through SpaceControl~\cite{fedele2025spacecontrol}.
The \textsc{aDSL} mesh serves as a spatial constraint for a pretrained generator such as Trellis~\cite{Xiang2024}, improving geometric detail and surface appearance while preserving global structure, as illustrated in \Cref{fig:high-fidelity}.
This retains program editability and semantic organization while delegating high-frequency detail to the external model. \looseness=-1

\paragraph{Scene Generation}
\textsc{aDSL} supports scenes by placing multiple objects in a shared coordinate frame.
Its hierarchy separates object definitions from scene layout: objects are independent semantic sub-programs, while the scene level specifies placement constraints and inter-object relations.
This structure lets objects or sub-structures be exported, refined by an external generator, and recomposed under the original constraints.
As shown in \cref{fig:scene-composition}, the resulting scenes gain rich visual detail while keeping global structure explicit, controllable, and editable.\looseness=-1

\paragraph{Interactive User Integration.}

Our system can be embedded in a chat-style front-end for iterative 3D creation and editing.
As shown in \cref{fig:openclaw}, a user issues an open-ended natural-language request, the agent translates it into detailed object attributes, refines the program over multiple rounds, and returns a preview for inspection.
Users can then continue the conversation to request edits or regeneration, making the creation process interactive and controllable. \looseness=-1

\section{Conclusion} \label{sec:conclusion}

In this paper, we presented a training-free framework for agentic 3D creation through the joint design of an agent-centric DSL and a role-specialized multi-agent system.
By representing 3D assets as executable, structured programs, our approach bridges high-level semantic intent and low-level geometry.
The core contribution is this joint design: a compositional DSL with spatial reasoning operators, tightly coupled with a Plan--Execute--Critic loop for iterative generation, verification, and repair of 3D programs.
Our evaluation shows that this co-design improves robustness, controllability, and editability, while naturally supporting downstream applications such as articulated asset modeling, shape editing, and scene-level composition. \looseness=-1

Our current system still has several limitations and clear directions for future work.
First, final output quality remains bounded by the expressiveness of the DSL and its geometric primitives; highly complex geometry, appearance, and material effects may require tighter integration with learned high-fidelity generators.
Second, although the \emph{Critic} provides useful feedback for iterative repair, its verification is still largely based on 2D renderings and may suffer from perspective ambiguity.
Third, the framework currently relies on strong proprietary LLMs for reliable long-horizon spatial reasoning and repair, limiting its accessibility.
Distilling these capabilities into open-source language models is an important next step toward making agentic 3D content creation more broadly accessible. \looseness=-1


\bibliographystyle{ACM-Reference-Format}
\bibliography{ref/reference} 

\clearpage
\appendix
\section{Experimental Details}
\label{sec:appendix_experimental_details}

\paragraph{Implementation Details}
We use Gemini 3 Pro~\cite{geminiteam2025gemini3} with temperature $1.0$ as the backbone model for all agents.
The self-correction loop runs for at most $R=10$ rounds and stops early when the \emph{Critic} reports no actionable issues.
Although our workflow supports optional user feedback, all reported experiments are fully automatic and do not use user feedback during generation or refinement.
For visual critique, we render eight views per shape at $45^\circ$ azimuth intervals and a fixed $15^\circ$ elevation.
All images are rendered at $1024 \times 1024$ resolution with neutral materials and environment lighting.
To ensure a fair comparison, we use the models with similar capacity for baselines that require LLM integration: Gemini 3 Pro~\cite{geminiteam2025gemini3} for Scene Language and ShapeCraft, and Claude Opus 4.5~\cite{anthropic2025claudeopus45} for BlenderMCP.
We standardize the rendering protocol across all methods instead of relying on each baseline's native renderer.

\paragraph{Metrics}
We evaluate generated shapes with four complementary metrics to capture the semantic, visual, and structural quality.
To penalize invalid outputs, we assign a score of zero to all metrics whenever a method fails to produce a valid output.
\begin{itemize}[leftmargin=*,itemsep=2pt]
\item[-]\emph{CLIP-Score}~\cite{radford2021learning} measures global semantic alignment as the cosine similarity between the input embedding and multi-view renderings of the generated shape.
\item[-]\emph{VQAScore}~\cite{lin2024evaluating} estimates whether the rendered shape visually entails the input text with CLIP-FlanT5; we use it only for text-to-shape generation.
\item[-]\emph{FID}~\cite{heusel2017gans}
measures similarity between generated and ground-truth shapes from rendered views. We extract  features with Inception-v3~\cite{szegedy2016rethinking} and average FID across canonical views; we use it for image-to-shape generation.
\item[-]\rh{\emph{Execution Success Rate} reports the fraction of prompts for which a method completes and produces a valid renderable mesh.}
\end{itemize}








\section{DSL Definition}
\label{sec:dsl_definition}

\begin{table*}[t]
    \centering
    \caption{Public types and operator families of the relational 3D programming interface. Signed axes follow the world convention $+x$ right, $+y$ inward, and $+z$ up.}
    \label{tab:dsl_grammar}
    \setlength{\tabcolsep}{5pt}
    \renewcommand{\arraystretch}{1.08}
    \small
    \begin{tabularx}{\textwidth}{@{}>{\raggedright\arraybackslash}p{0.28\textwidth} X@{}}
        \toprule
        \textbf{Type or operator family} & \textbf{Public interface and semantics}\\
        \midrule
        \texttt{P}, \texttt{T}, \texttt{Asset} & 3D vectors, $4\times4$ transforms, and hierarchical geometry containers with uniquely named direct parts.\\
        \texttt{attach\_part}, \texttt{detach\_part}, \texttt{copy}, \texttt{concat\_shapes} & Construct, revise, copy, and combine explicit asset hierarchies.\\
        \texttt{Cube/cube}, \texttt{Sphere/sphere}, \texttt{Cylinder/cylinder} & Equivalent capitalized and lowercase primitive constructors; cylinders accept either endpoints or a height and cardinal axis.\\
        \texttt{boolean\_union}, \texttt{boolean\_intersection}, \texttt{boolean\_difference}, \texttt{boolean\_xor} & Constructive solid geometry operators that retain operand hierarchy.\\
        \texttt{translation\_matrix}, \texttt{scaling\_matrix}, \texttt{rotation\_matrix} & Matrix constructors; positive axis--angle rotations obey the right-hand rule.\\
        \texttt{transform\_shape}, \texttt{translate\_shape}, \texttt{scale\_shape}, \texttt{rotate\_shape} & Return transformed copies; rotation supports axis--angle and Euler forms.\\
        \texttt{shape\_aabb}, \texttt{shape\_min}, \texttt{shape\_max}, \texttt{shape\_size}, \texttt{shape\_center} & Query world-space axis-aligned bounds.\\
        \texttt{shape\_anchor} & Query named AABB faces, edges, and corners such as \texttt{top} and \texttt{left\_front\_top}.\\
        \texttt{shape\_support}, \texttt{shape\_bounds\_along}, \texttt{shape\_extent\_along} & Query support points and extents along arbitrary directions, including rotated-contact reasoning.\\
        \texttt{align\_centers}, \texttt{align\_anchors}, \texttt{place\_on\_axis}, \texttt{offset\_from} & Express relational placement by centers, named anchors, signed target surfaces, or offsets.\\
        \texttt{distribute\_along\_axis}, \texttt{stack\_shapes} & Linear repeated layouts with center spacing or boundary gaps; the first input remains fixed.\\
        \texttt{grid\_shapes}, \texttt{radial\_shapes} & Grid and radial layouts with explicit plane/axis conventions and optional instance rotation.\\
        \texttt{Asset.revolute}, \texttt{Asset.prismatic}, \texttt{Asset.fixed} & Ergonomic joints whose axes are expressed in the zero-pose joint/child frame.\\
        \texttt{Asset.attach\_joint} & Low-level joint attachment; unlike the ergonomic methods, its axis is expressed in the parent-link frame.\\
        \bottomrule
    \end{tabularx}
\end{table*}

We summarize the public types and operators of our DSL in \cref{tab:dsl_grammar}.
The coordinate system uses $+x$ to the right, $+y$ inward, and $+z$ upward.
Transforms and layout operators return new asset trees, whereas hierarchy, joint, and appearance methods mutate the receiving \texttt{Asset}.
For articulated assets, child geometry is first placed in the parent link's zero-pose coordinates and is then rebased into the joint frame; revolute signs follow the right-hand rule.
The following listing reproduces the complete DSL reference supplied to the agents, followed by the in-context modeling example.

\noindent {Complete aDSL modeling reference:}
\begin{lstlisting}
# aDSL modeling reference

Import the public API from `adsl.core`:

```python
from adsl.core import *
```

The world coordinate convention is `+x` right, `+y` inward, and `+z` up.
Lengths use caller-defined scene units. Rotation angles passed to
`rotation_matrix` and the axis/angle form of `rotate_shape` are degrees. Joint
positions use radians for revolute joints and scene units for prismatic joints.
All positive axis-angle rotations follow the right-hand rule. For a quick sign
check, a positive 90-degree rotation maps `+y` toward `+z` around `+x`, `+z`
toward `+x` around `+y`, and `+x` toward `+y` around `+z`. Negating the axis
reverses the positive rotation direction.

Transforms and layout functions return new Asset trees. `attach_part`, joint
methods, and appearance setters modify the receiving Asset and return an Asset
for fluent construction.

## Assets and hierarchy

```python
Asset(label: str = "Asset")
Asset.attach_part(name: str, shape: Asset) -> Asset
Asset.detach_part(name: str) -> None
Asset.copy() -> Asset
concat_shapes(shapes: Iterable[Asset], *, label: str | None = None) -> Asset
```

`attach_part` records a named modeling subpart and preserves the hierarchy.
Names must be unique among the direct children of one parent.

`concat_shapes` returns a container whose children are named `part_0`,
`part_1`, and so on. Use explicit `attach_part` calls when semantic names matter.

Example:

```python
desk = Asset("desk")
desk.attach_part("desktop", Cube((1.4, 0.7, 0.06), center=(0, 0, 0.73)))
legs = Asset("legs")
for index, position in enumerate(((-0.6, -0.25), (-0.6, 0.25), (0.6, -0.25), (0.6, 0.25)), 1):
    legs.attach_part(
        f"leg_{index}",
        Cube((0.06, 0.06, 0.7), center=(position[0], position[1], 0.35)),
    )
desk.attach_part("legs", legs)
scene = desk
```

## Primitives

The capitalized constructors and lowercase constructors are equivalent public
forms. A scalar cube scale creates equal x/y/z dimensions.

```python
Cube(scale: float | Sequence[float], center=(0, 0, 0), color=(1, 1, 1), alpha=None) -> Asset
Sphere(radius: float, center=(0, 0, 0), color=(1, 1, 1), alpha=None) -> Asset
Cylinder(
    radius: float,
    p0: Sequence[float] | None = None,
    p1: Sequence[float] | None = None,
    *,
    height: float | None = None,
    center=(0, 0, 0),
    axis="z",
    color=(1, 1, 1),
    alpha=None,
) -> Asset
cube(...), sphere(...), cylinder(...)
```

A cylinder requires either both endpoints `p0`/`p1`, or `height` with a
cardinal `axis`. Endpoints are the centers of the circular end caps. A cylinder
is symmetric along its length, so negating `axis` only swaps which end is
considered `p0` versus `p1`; it does not change the visible geometry.

## Boolean operations

```python
boolean_union(*shapes: Asset) -> Asset
boolean_intersection(*shapes: Asset) -> Asset
boolean_difference(base: Asset, *subtractors: Asset) -> Asset
boolean_xor(*shapes: Asset) -> Asset
```

Boolean results retain their operand hierarchy for inspection.

## Transformations

```python
translation_matrix(offset: Sequence[float]) -> T
scaling_matrix(scale: float | Sequence[float], center=(0, 0, 0)) -> T
rotation_matrix(axis: str | Sequence[float], angle: float, center=(0, 0, 0)) -> T
transform_shape(shape: Asset, matrix: T) -> Asset
translate_shape(shape: Asset, offset: Sequence[float]) -> Asset
scale_shape(shape: Asset, scale: float | Sequence[float], center=None) -> Asset
rotate_shape(shape: Asset, axis, angle: float, center=None) -> Asset
rotate_shape(shape: Asset, *, euler: Sequence[float], center=None) -> Asset
```

Signed cardinal axes are `+x`, `-x`, `+y`, `-y`, `+z`, and `-z`; bare axis
letters mean their positive direction. Axis-angle rotation follows the
right-hand rule described above. The `euler=(x, y, z)` form accepts degrees and
applies the x rotation first, then y, then z (combined matrix `Rz @ Ry @ Rx`).
When a transform center is omitted, `scale_shape` and `rotate_shape` use the
current AABB center.

## Bounds and anchors

```python
shape_aabb(shape: Asset) -> tuple[P, P]
shape_min(shape: Asset) -> P
shape_max(shape: Asset) -> P
shape_size(shape: Asset) -> P
shape_center(shape: Asset) -> P
shape_anchor(shape: Asset, anchor: str = "center") -> P
shape_support(shape: Asset, direction: str | Sequence[float]) -> P
shape_bounds_along(shape: Asset, direction) -> tuple[float, float]
shape_extent_along(shape: Asset, direction) -> float
```

The first six functions use world-space axis-aligned bounds. Anchor tokens map
to AABB sides:

- `left` / `right`: minimum / maximum x
- `front` / `back`: minimum / maximum y
- `bottom` / `top`: minimum / maximum z

Unspecified axes use the center. For example, `top` is the center of the top
face and `left_front_top` is a corner. Support and directional-bound functions
should be used for arbitrary directions and rotated contact reasoning.

## Alignment and placement

```python
align_centers(shape: Asset, target: Asset, axes=("x", "y", "z")) -> Asset
align_anchors(
    shape: Asset,
    target: Asset | Sequence[float],
    anchor: str = "center",
    target_anchor: str | None = None,
    offset=(0, 0, 0),
) -> Asset
place_on_axis(shape: Asset, target: Asset | float, axis="+z", gap=0.0) -> Asset
offset_from(
    shape: Asset,
    reference: Asset | Sequence[float | None] | None,
    offset: Sequence[float | None],
) -> Asset
```

`align_anchors` aligns one source AABB anchor with an Asset anchor or an exact
world point. `target_anchor` is valid only for an Asset target and defaults to
the same name as `anchor`.

`place_on_axis` uses the axis sign to choose direction. For `+z`, the source
bottom is placed above the target top. For `-z`, the source top is placed below
the target bottom. A numeric target is the boundary coordinate. `gap` must be
non-negative.

`offset_from` positions selected center coordinates relative to an Asset center,
a point, or the origin. A `None` coordinate leaves that source coordinate
unchanged.

## Repeated layouts

```python
distribute_along_axis(shapes, axis="+x", spacing=1.0) -> Asset
stack_shapes(shapes, axis="+z", gap=0.0) -> Asset
grid_shapes(
    shapes,
    rows=None,
    cols=None,
    spacing=(1.0, 1.0),
    plane="xy",
    center=(0, 0, 0),
    order="row-major",
) -> Asset
radial_shapes(
    shapes,
    radius,
    axis="+z",
    center=(0, 0, 0),
    start_angle=0.0,
    sweep=360.0,
    *,
    rotate_with_layout=False,
    rotation_offset=0.0,
) -> Asset
```

For `distribute_along_axis` and `stack_shapes`, the **first input shape is the
fixed base** and remains at its original center. Later shapes are placed in
sequence along the signed axis. Distribution uses center-to-center `spacing`;
stacking uses `gap` between neighboring AABB boundaries. Both distances must be
non-negative.

`grid_shapes` centers the complete grid at `center`. `spacing` is the
center-to-center pitch in the two axes named by `plane`. Columns increase along
the positive first plane axis; rows increase along the negative second plane
axis. For `plane="xy"`, columns run along `+x`; the first row is on the `+y`
side, and later rows advance toward `-y`.

`radial_shapes` uses evenly spaced slots without duplicating the first slot for
a 360-degree sweep. Partial arcs include both endpoints. With
`rotate_with_layout=False`, input orientations are unchanged. With
`rotate_with_layout=True`, every shape is first rotated around its own center by
its slot angle plus `rotation_offset`, then translated. The input orientation at
zero degrees is the pattern reference. Positive slot angles follow the
right-hand rule around the signed `axis`; negating `axis` reverses the sweep.
The zero-angle radial direction is `+x` for a z axis, `+y` for an x axis, and
`+z` for a y axis. For example, around `axis="+z"`, zero degrees lies on `+x`
and positive angles sweep toward `+y`.

Bicycle-spoke example:

```python
spokes = radial_shapes(
    [Cube((0.45, 0.015, 0.015)) for _ in range(12)],
    radius=0.225,
    axis="+z",
    rotate_with_layout=True,
)
```

## Articulation

```python
Asset.revolute(
    child: Asset | str,
    *, axis=(0, 0, 1), limit=(-pi, pi), origin=(0, 0, 0),
    initial=0.0, joint_name=None, effort=None, velocity=None,
) -> Asset
Asset.prismatic(
    child: Asset | str,
    *, axis=(0, 0, 1), limit=(0, 1), origin=None, initial=0.0,
    towards=None, joint_name=None, effort=None, velocity=None,
) -> Asset
Asset.fixed(child: Asset | str, *, joint_name=None, origin=None) -> Asset
Asset.attach_joint(
    joint_name: str,
    child_link: Asset,
    *, joint_type="revolute", axis=(0, 0, 1), origin=None,
    limit=None, initial=0.0, effort=None, velocity=None,
) -> Asset
```

The ergonomic `revolute`, `prismatic`, and `fixed` methods accept the name of an
existing direct part or an Asset. When an Asset is passed directly, `joint_name`
is required. Place the child geometry in the parent's zero-pose coordinates
before creating the joint. The method rebases the child by `inverse(origin)`
into the joint frame, so `origin` is the hinge/pivot/slide frame expressed in
the parent link. A point origin supplies translation only; a 4x4 origin may also
rotate the joint frame.

For the ergonomic methods, `axis` is expressed in the joint/child frame at the
zero pose. Positive revolute motion follows the right-hand rule around that
axis; positive prismatic motion translates along the axis. Negating the axis
reverses the meaning of positive joint values. Choose `axis`, signed `limit`,
and `initial` together so the initial and endpoint poses move the part in the
intended physical direction. The pivot location alone does not determine which
way a lid or door opens.

`attach_joint` is the low-level exception: its `axis` is expressed directly in
the **parent-link frame**, not the child/joint frame. Prefer the ergonomic
methods unless that distinction is intentional.

For a freely rotating revolute joint, use `limit=None` or
`limit=(-float("inf"), float("inf"))`.

`initial` must be finite and within a finite `limit`. `towards` on a prismatic joint may
name a direct sibling, provide an Asset, provide a point, or select the parent
origin; it flips the axis when necessary and raises if the target cannot be
resolved. It chooses the positive slide direction only; limits and initial
values remain measured along that resolved direction.

Before finalizing an articulated object, reason about both the zero pose and at
least one nonzero pose. Example: if a closed laptop lid extends from an x-axis
hinge toward `-y`, then `axis="-x"` with positive limits rotates the lid toward
`+z`:

```python
laptop.attach_part("lid", lid_in_closed_parent_coordinates)
laptop.revolute(
    "lid", axis="-x", origin=hinge_point,
    limit=(0.0, 2.18), initial=1.83,
)
```

Using `axis="+x"` for the same geometry would require equivalent negative
limits and an initial value such as `-1.83`. If the lid extends toward `+y`
instead, reverse these signs.
\end{lstlisting}

\noindent {DSL example prompt template:}
\begin{lstlisting}
```python
from adsl.core import *
import numpy as np


class Book(Asset):
    def __init__(self, scale: P):
        super().__init__(label="Book")
        self.body = self.attach_part(
            "body",
            cube(scale, color=(0.6, 0.3, 0.1), alpha=0.8),
        )


class Books(Asset):
    def __init__(self, width: float, length: float, book_height: float, num_books: int):
        super().__init__(label="Books")
        rng = np.random.default_rng(7)

        def make_book() -> Asset:
            book = Book(scale=(width, length, book_height))
            book = translate_shape(
                book,
                (
                    rng.uniform(-0.05, 0.05),
                    rng.uniform(-0.05, 0.05),
                    0,
                ),
            )
            angle_degrees = rng.uniform(-15.0, 15.0)
            return rotate_shape(book, axis="+z", angle=angle_degrees)

        self.stack = self.attach_part(
            "stack",
            stack_shapes([make_book() for _ in range(num_books)], axis="z"),
        )


class Table(Asset):
    def __init__(self, top_scale: P, leg_scale: P):
        super().__init__(label="Table")

        # Put the feet on z=0 and support the tabletop at the tops of the legs.
        tabletop_center_z = leg_scale[2] + top_scale[2] / 2.0
        tabletop = cube(
            top_scale,
            center=(0.0, 0.0, tabletop_center_z),
            color=(0.4, 0.2, 0.1),
        )
        self.tabletop = self.attach_part("tabletop", tabletop)

        leg_alignments = (
            ("left_front_top", "left_front_bottom"),
            ("right_front_top", "right_front_bottom"),
            ("right_back_top", "right_back_bottom"),
            ("left_back_top", "left_back_bottom"),
        )
        for index, (leg_anchor, tabletop_anchor) in enumerate(leg_alignments, 1):
            leg = cube(leg_scale, color=(0.3, 0.15, 0.07))
            leg = align_anchors(
                leg,
                tabletop,
                anchor=leg_anchor,
                target_anchor=tabletop_anchor,
            )
            self.attach_part(f"leg_{index}", leg)


class TableWithBooks(Asset):
    def __init__(self):
        super().__init__(label="TableWithBooks")
        table = Table(top_scale=(1.0, 0.6, 0.05), leg_scale=(0.08, 0.08, 0.70))
        self.table = self.attach_part("table", table)

        books = Books(width=0.21, length=0.29, book_height=0.05, num_books=3)
        books = align_anchors(
            books,
            table.tabletop,
            anchor="bottom",
            target_anchor="top",
        )
        self.books = self.attach_part("books", books)


scene = TableWithBooks()
```
\end{lstlisting}



\section{Prompt Templates}
\label{sec:prompt_templates}

In this section, we provide the prompt templates used by the agent workflow.
At runtime, \texttt{[DSL\_DOC]} and \texttt{[DSL\_EXAMPLE]} are replaced by the material in \cref{sec:dsl_definition}; the articulation-guidance placeholders are instantiated only for tasks that expose articulation APIs.

\noindent {Prompt for \emph{Planner}:}
\begin{lstlisting}
You are a planner in a 3D modeling workflow. Your task is to analyze the user's instruction and parse it into a structured format for the Coder and Critic to work on. You should carefully analyze the description, extracting as much valuable and precise information for the modeler to refer to.

Your structured output must contain:
- `object_name`: the modeled object's name
- `components`: named components with precise descriptions
- `relations`: spatial, structural, functional, and articulation relations
- `critic_checklist`: verifiable and precise review rules

Here is the Domain Specific Language that will be used during the entire process. Your plan and recommendation should strictly follow the principles that they can be satisfied by the provided functions. [ARTICULATION_PLANNER_GUIDANCE]

[DSL_DOC]
\end{lstlisting}

\noindent {Prompt for \emph{Debugger}:}
\begin{lstlisting}
You are a code debugger. You will receive an execution error for a 3D modeling program. Read exactly the workspace-relative `assigned_source` supplied in the input before diagnosing it; do not infer another filename. Your task is to identify the bug in the code and provide suggested fixes.

The code is meant for 3D modeling using a domain-specific language in Python. Ensure that your suggestions adhere to the syntax and semantics of this modeling language. Do not edit the file.

The DSL documentation is as follows:
[DSL_DOC]

Here is an example of modeling a scene with aDSL:
[DSL_EXAMPLE]

Return the structured `bug_description` and `suggested_fix` fields.
\end{lstlisting}

\noindent {Prompt for \emph{Coder}:}
\begin{lstlisting}
You are a Coder. Write 3D modeling code using the provided aDSL Domain-Specific Language (DSL), according to user requirements or review feedback.

[DSL_DOC]

Here is an example of modeling a scene with aDSL:
[DSL_EXAMPLE]
IMPORTANT: THE CLASSES ABOVE ARE JUST EXAMPLES, YOU CANNOT USE THEM IN YOUR PROGRAM!

STRICTLY follow these rules:
1. Only use the functions, classes, and imported libraries exposed by `from adsl import *`. For a new asset, use `write_file` exactly once to write the complete assigned program. For a correction, first use `read_file`, then use one or more exact `apply_patch` calls. Never return source code in the assistant response.
2. Define reusable components as subclasses of `Asset` to structure your code.
3. Build geometry with the documented primitives such as `Cube`, `Sphere`, and `Cylinder`.
4. You should STRICTLY follow the coordinate system: +x is right, +y is inward (into the screen), +z is up.
5. Prefer the spatial reasoning helpers to express positions and relationships explicitly: `place_on_axis`, `align_centers`, `align_anchors`, `offset_from`, `distribute_along_axis`, `grid_shapes`, `radial_shapes`, `stack_shapes`, `translate_shape`, `rotate_shape`, the AABB query helpers `shape_center`, `shape_min`, `shape_max`, `shape_size`, `shape_aabb`, `shape_anchor`, and the directional query helpers `shape_support`, `shape_bounds_along`, `shape_extent_along`. Use anchor names like `top`, `front`, or `left_front_top` when placing shapes by faces, edges, or corners. Use `grid_shapes(...)` or `radial_shapes(...)` for repeated arrays instead of manual placement loops when they match the layout. When radial instances should rotate with their slots, use `radial_shapes(..., rotate_with_layout=True)`; otherwise their input orientations remain unchanged. Use directional queries when reasoning about rotated parts or span along arbitrary directions. When one face/edge/corner relationship determines the full placement, prefer one `align_anchors(...)` call instead of chaining separate axis moves.
6. [ARTICULATION_CODER_GUIDANCE]
7. Use boolean operations to model complex geometry.
8. Finish by assigning the final `Asset` to a variable named `scene`.

You should be creative and precise.
\end{lstlisting}

\noindent {Prompt for \emph{Image Critic}:}
\begin{lstlisting}
You are a Critic. You need to find issues in the provided rendered images based on the user requirement and the planner's checklist. In addition, you will be told the maximum allowed number of refinement interaction rounds and the current round you are in; you must decide how to prioritize existing problems based on the current round.

**IMPORTANT**: If you receive conclusions from the Code Critic, treat them as authoritative. When your visual impression conflicts with the Code Critic's conclusion, defer to the Code Critic and do not request code changes based on the renders.

The images are rendered from different views. In the first image, the coordinate system is as follows: +x is right, +y is into the screen, +z is up. The following seven views are rotated around the vertical (z) axis counter-clockwise by 45 degrees, 90 degrees, 135 degrees, 180 degrees, 225 degrees, 270 degrees, and 315 degrees respectively.

You should follow these principles when reviewing:
1. Focus on the most critical issues that impact correctness and functionality.
2. Offer clear suggestions for fixes rather than just pointing out problems.
3. Address only **ONE** most critical issue if multiple are found.
4. Your suggestions must be consistent with previous critic comments in earlier rounds to ensure coherence throughout the refinement process.
5. Focus on major issues that affect the overall structure, functionality, and spatial relationships.
6. Use **ALL** views together to understand the full 3D structure. If a component is occluded in one view, infer its presence, absence, and placement from other views.

Return structured output with `approved`, `observations`, and `required_changes`. Set `approved=true` only when the judgement is APPROVED; for REVISION_NEEDED, put the single most critical actionable change in `required_changes`.
\end{lstlisting}

\noindent {Prompt for \emph{Code Critic}:}
\begin{lstlisting}
You are a strict Critic. Your job is to reconcile the Coder's DSL with the Image Critic's feedback, and then give actionable feedback to both.

## Inputs you will receive

1. The user requirement for the 3D scene.
2. The Coder's program, available through the assigned `read_file` tool.
3. The images rendered from different views.
4. The suggestions from the Image Critic that reviewed the rendered images.

## Your tasks

1. Read exactly the workspace-relative `assigned_source` supplied in the input; do not infer another filename. Inspect the Coder's implementation with the Image Critic's suggestions.
2. Decide whether the Image Critic's suggestions are valid **based on the code**.
3. For valid suggestions, provide feedback to the Coder for necessary revisions.
4. For invalid suggestions, provide feedback to the Image Critic to clarify misunderstandings.

## CORE RULE

- Use the Coder's answer as the primary reference to judge the validity of the Image Critic's suggestions.
- You MUST TRUST THE CODE LOGIC to avoid potential misunderstanding and visual artifacts from rendered images.
- Never hedge by saying the Image Critic "might be wrong" while also telling the Coder to "double-check" the same point. Pick one side based on the DSL and commit.
- When articulation APIs are available, judge joint placement using the DSL frame contract: before `revolute`, `prismatic`, or `fixed`, the moving child must already be placed in the parent link's zero-pose coordinates. The joint call rebases the child by `inverse(origin)`, so evaluate the post-joint child link frame, not only the child's local pre-joint construction. If a child is built around local `(0, 0, 0)` and passed with a nonzero `origin` without first being aligned into the parent zero-pose location, flag it as a frame error.
- Also verify motion direction, not just pivot location. Positive revolute values follow the right-hand rule around the ergonomic method's joint/child-frame axis, while positive prismatic values move along that axis. Evaluate the signed `axis`, `limit`, and `initial` together at a representative nonzero pose; flag joints whose configured motion sends a lid, door, lever, or similar part through the body or opposite its intended direction. Low-level `attach_joint(...)` is the exception whose axis is in the parent-link frame.

## APPROVAL CONTRACT

- `approved` evaluates whether the Coder's current implementation and rendered result satisfy the user requirement and may end the refinement loop. It does **not** indicate whether you agree with the Image Critic.
- Set `approved=true` only when no valid revision remains and `required_changes` is empty.
- If any Image Critic suggestion is valid, set `approved=false` and put every valid, actionable revision in `required_changes`.
- Put invalid Image Critic suggestions in `image_critic_corrections`. Rejecting an invalid suggestion does not by itself require a Coder revision.
- Never return `approved=true` together with a non-empty `required_changes` list.

Return structured output with `approved`, `observations`, `required_changes`, and `image_critic_corrections`. Valid suggestions belong in `required_changes`; misunderstandings belong in `image_critic_corrections`.

## DSL Reference

[DSL_DOC]

Here is an example of modeling a scene with the DSL:
[DSL_EXAMPLE]
\end{lstlisting}


\end{document}